\documentclass[a4paper,pra,amsmath,amssymb,floatfix,longbibliography,aps,twocolumn,superscriptaddress,accepted=2023-03-31]{quantumarticle}
\pdfoutput=1
\usepackage{amsfonts,color,physics,enumerate}
\usepackage{hyperref}
\usepackage{dsfont}
\usepackage{latexsym} 
\usepackage{mathtools}
\usepackage{tabularx}

\usepackage[utf8]{inputenc}
\usepackage[english]{babel}
\usepackage[T1]{fontenc}

\usepackage[numbers,sort&compress]{natbib}

\newcommand{\eave}[1]{\mathbb{E} \left[ #1 \right]}

\newcommand{\RR}{\mathbb{R}}

\newcommand{\sopvec}[1]{\bar{\bar{ #1 }}}

\begin{document}
\title{Universal equilibration dynamics of the Sachdev-Ye-Kitaev model}
\author{Soumik Bandyopadhyay}
\thanks{These two authors contributed equally.}
\affiliation{Pitaevskii BEC Center, CNR-INO and Dipartimento di Fisica, 
             Università di Trento, Via Sommarive 14, Trento, I-38123, Italy}
\author{Philipp Uhrich}
\thanks{These two authors contributed equally.}
\affiliation{Pitaevskii BEC Center, CNR-INO and Dipartimento di Fisica, 
             Università di Trento, Via Sommarive 14, Trento, I-38123, Italy}
\author{Alessio Paviglianiti}
\affiliation{Pitaevskii BEC Center, CNR-INO and Dipartimento di Fisica, 
             Università di Trento, Via Sommarive 14, Trento, I-38123, Italy}
\affiliation{International School for Advanced Studies (SISSA), 
             via Bonomea 265, 34136 Trieste, Italy}
\author{Philipp Hauke}
\affiliation{Pitaevskii BEC Center, CNR-INO and Dipartimento di Fisica, 
             Università di Trento, Via Sommarive 14, Trento, I-38123, Italy}

\begin{abstract}
		Equilibrium quantum many-body systems in the vicinity of phase 
		transitions generically manifest universality. 
		In contrast, limited knowledge has been gained on possible universal characteristics in the non-equilibrium evolution of systems in quantum critical phases.
        In this context, universality is generically attributed to the 
        insensitivity of observables to the microscopic system parameters and 
        initial conditions.
		Here, we present such a universal feature in the equilibration dynamics of the Sachdev-Ye-Kitaev (SYK) Hamiltonian---a paradigmatic system of disordered, all-to-all 
		interacting fermions that has been designed as a phenomenological description of quantum critical regions. 
		We drive the system far away from equilibrium by performing a global quench, and track how its ensemble average relaxes to a steady state.
		Employing state-of-the-art numerical simulations for the exact 
		evolution, we reveal that the disorder-averaged evolution of few-body 
		observables, including the quantum Fisher information and 
		low-order moments of local operators, exhibit within numerical 
		resolution a universal equilibration process. 
		Under a straightforward rescaling, data that correspond to different 
		initial states collapse onto a universal curve, which can be well 
		approximated by a Gaussian throughout large parts of the evolution. 
		To reveal the physics behind this process, we formulate a general theoretical framework based on the Novikov--Furutsu theorem. This framework extracts the disorder-averaged dynamics of a many-body system as an effective dissipative evolution, and can have applications beyond this work. 
		The exact non-Markovian evolution of the SYK ensemble is very well 
		captured by Bourret--Markov approximations, which contrary to common 
		lore become justified thanks to the extreme chaoticity of the system, 
		and universality is revealed in a spectral analysis of the 
		corresponding Liouvillian. 
\end{abstract}
	
\maketitle
	
\section{Introduction}\label{s:intro}

Whether and how a perturbed system equilibrates have been fundamental issues of 
statistical mechanics since the laying of its foundation. One question that has intrigued 
researchers for almost a century is the thermalization of an isolated quantum system under
its unitary evolution~\cite{vonNeumann29, polkovnikov_2011, eisert_2015, GogolinEisert16}.
In the last two decades, this process has experienced a revitalized surge of interest, 
thanks to experimental breakthroughs in realizations of synthetic many-body 
systems~\cite{lewenstein_book, bloch_2012, blatt_2012, hauke_2012, 
georgescu_2014, gross_2017, altman_2021}.
An unprecedented control over system parameters now enables laboratory 
investigations using
quantum systems in almost ideal, isolated 
conditions~\cite{strohmaier_2010, trotzky_2012, gring_2012, langen_2013, 
jurcevic_2014, smith_2016, kaufman_2016, neill_2016, clos_2016, neyenhuis_2017, liu_2018, 
tang_2018, kim_2018, prufer_2018, zhou_2021}.
On the theory side, a main obstacle for arriving at a unified understanding of 
out-of-equilibrium quantum dynamics is the absence of a universal principle 
that would be as general as the minimization of free energy for equilibrium 
phase transitions~\cite{nishimori_2010, sachdev_2011}.  Nevertheless, powerful 
frameworks have been developed to explain the thermalization of a quantum 
system, perhaps the most successful being the eigenstate thermalization 
hypothesis (ETH)~\cite{Deutsch91, Srednicki94, Rigol_etal08, DAlessio_etal16}. 
According to the ETH, for quantum chaotic systems, particularly for ergodic 
systems, thermalization and equilibration are tantamount, since under quantum 
chaotic dynamics a perturbed system equilibrates to a state that for local 
observables is indistinguishable from a Gibbs thermal state.  
A well accepted mechanism 
for thermalization in an isolated quantum system is 
that initially localized information is distributed among the system's degrees 
of freedom, and thus becomes irretrievable through any local operation at later 
times. This process, referred to as scrambling~\cite{lashkari_2013, 
hosur_2016}, 
occurs in quantum many-body lattice models~\cite{bohrdt_2017, iyoda_2018, 
bentsen_2019_1}, conformal field theories \cite{RobertsStanford2015}, and black 
holes
alike~\cite{hayden_2007, sekino_2008}, and has been experimentally probed in 
different physical systems~\cite{joshi_2020, blok_2021, zhu_2021}.

A central role in bridging the different paradigms of scrambling and chaotic 
dynamics has been taken by the Sachdev-Ye-Kitaev (SYK) 
model~\cite{sachdev_1993, sachdev_2015, Kitaev_talk, maldacena_2016_2, 
gu_2020}.
This model, which consists of disordered all-to-all interactions, 
was originally designed as a prototype for so-called strange 
metals~\cite{sachdev_1993, sachdev_2010_iop, sachdev_2015, song_2017}, and has 
been found to be holographically dual to black holes with 
two-dimensional anti-de Sitter horizons~\cite{sachdev_2010, sachdev_2015, 
maldacena_2016_2, davison_2017, kitaev_2018, sachdev_2019}.
Like the black holes~\cite{sekino_2008}, this model exhibits fast scrambling 
dynamics by saturating the upper bound of the quantum Lyapunov 
exponent~\cite{maldacena_2016_1, maldacena_2016_2}. In this sense, the SYK model 
is maximally chaotic, which has spurred much recent theoretical investigations 
into its chaotic~\cite{garcia_2016, cotler_2017, garcia_2018, numasawa_2019, 
winer_2020, kobrin_2021} and thermalization properties~\cite{magan_2016, 
sonner_2017, eberlein_2017, LouwKehrein_2022}
, as well as its post-quench dynamics \cite{Davidson_etal2017,Haldar_etal_2020,SamuiSorokhaibam_2021,Carrega_etal_2021,LarzulSchiro_2022}.
In addition, proposals for quantum simulating the SYK model on digital 
devices \cite{GarciaAlvarez_etal2017} and analog systems based on the solid 
state \cite{pikulin_2017, chew_2017, chen_2018} and ultracold 
atoms \cite{danshita_2017, WeiSedrakyan_2021} have been put forward.

Despite the recent progresses in understanding the dynamics of this 
paradigmatic model and of quantum many-body systems in general, it remains an 
outstanding challenge to extract universal quantum out-of-equilibrium 
behavior~\cite{marcuzzi_2014, marcuzzi_2015, trapin_2018, 
heyl_2018, erne_2018, surace_2020, prakash_2020, berdanier_phdthesis_2020}. 
For slow near-adiabatic sweeps across a critical region, the Kibble-Zurek 
mechanism~\cite{kibble_1976, zurek_1985} has provided deep insights, including universal 
scaling laws~\cite{zurek_1985, delCampo_2014}. Here, we are interested in 
violent quenches, where a significant amount of energy is instantaneously 
injected into the system. 
Excepting few situations, such as non-thermal fixed points~\cite{berges_2008,
orioli_2015, berges_2015, karl_2017, prufer_2018, chatrchyan_2020, gresista_2021}, 
much less is known about universality in such far-from-equilibrium situations.

In this paper, we identify a \emph{universal equilibration} in quench dynamics of the 
complex SYK model, revealed in state-of-the-art numerical calculations for the
exact dynamics and reproduced analytically through a master equation.
In particular, employing a highly optimized exact diagonalization method for 
systems comprising up to $N = 20$ complex fermionic modes, we study how a 
system initialized in an eigenstate of some other Hamiltonian 
equilibrates to a steady state following a sudden global quench into the SYK 
model.
For a broad variety of few-body observables, including multipartite entanglement 
as given by the quantum Fisher information (QFI), the disorder-averaged evolution 
collapses onto a single curve after a simple amplitude rescaling, independent of (generic)
initial states. Over vast stretches of the dynamical evolution, this universal 
curve is well approximated by a Gaussian, with a fast decay on the order of the
time-scales of leading-order processes. 
Such a universality over the entire evolution goes significantly beyond what is observed in conventional equilibrating systems, where universal behavior independent of initial conditions can only be expected once the system reaches a 
fixed point of the dynamics (the final steady 
state~\cite{eisert_2015, GogolinEisert16, DAlessio_etal16} or a non-thermal fixed 
point~\cite{berges_2008, orioli_2015, berges_2015, karl_2017, prufer_2018, 
chatrchyan_2020, gresista_2021}). Thus, our findings may stimulate future investigations 
in non-equilibrium quantum many-body dynamics in order to identify similar universal 
dynamics in other models.

We substantiate our numerical findings by devising a Lindblad master equation 
(ME) that 
describes the Hamiltonian disorder average as an effective 
nonunitary time evolution. 
In this formalism, the unitary but disordered closed-system dynamics generated 
by the SYK model is mapped to one of a clean but dissipative system.
A detailed prescription for Hamiltonian disorder averaging has been introduced 
by Kropf et al. based on a matrix 
formalism \cite{andersson_2007, hall_2014, Kropf_etal16}. Here, we present an alternative, mathematically elegant route to Gaussian disorder 
averaging based on the Novikov--Furutsu theorem. 
In earlier works, this theorem has been applied in the context of averaging 
noise with finite correlation times in, for instance, quantum walks subjected to pure 
dephasing noise~\cite{Montiel_etal19, Ancheyta_etal2021}, stochastic Schr\"{o}dinger 
equations~\cite{Benatti_etal2012}, or proposals for simulating dissipation via noisy 
unitary dynamics~\cite{Chenu_etal2017, budini_2000, budini_2001, 
mildenberger_masterthesis2019}. 
We exploit that framework by formally promoting the quenched disorder to noise with 
infinite correlation time, permitting a fruitful application to a generic system with 
Gaussian disorder. 
To render the ensuing exact equations tractable, we employ decorrelation and 
Markovian approximations.
In contrast to standard lore, which seemingly preempts their use for the 
infinite correlation time of quenched disorder \cite{Visscher1974, 
Schekochihin01, Kubo57, vanVelsen1977, Kubo1985, vanVliet1988, Goderis1991}, 
these approximations capture the true quantum dynamics well, thanks to the 
extreme chaoticity of the SYK model. 
The resulting master equation successfully describes the super-exponential 
aspect of the equilibration process as well as the equilibrated steady state.
Even more, a spectral analysis of the associated Liouvillian provides valuable 
insights 
into the universal dynamics of few-body observables. 
As these results show, the master equation for the ensemble averaged state 
elegantly describes complex features of disordered quantum dynamics, and may 
thus constitute a powerful tool beyond the immediate context of this 
work.

The rest of this manuscript is organized as follows. 
In Sec.~\ref{s:quench_protocol}, we elaborate on a global quench protocol, employed to 
induce the equilibration dynamics. Then, in Sec.~\ref{s:unidynqfi}, we discuss the 
universality and super-exponential decay observed in the disorder ensemble 
averaged  
dynamics of the QFI. Sec.~\ref{s:master_eq} presents the formalism used to 
obtain the master 
equation, and further illustrates that the latter captures the salient features of the 
equilibration process. These sections constitute the main results of our study.
To provide a further in-depth analysis, we extend our study in Sec.~\ref{s:uni_dynSYK4} 
to the dynamics of operator moments, and explain the observed  universality in the 
equilibration process based on a spectral analysis of the Liouvillian.
In Sec.~\ref{s:conclusion}, we conclude the key findings of our study, and emphasize 
possible extensions and potential applications of the presented formalism. The main 
article is complemented by appendices that provide further details on the model, master 
equation formalism, as well as additional numerical results.

\begin{figure}[ht!]
 \includegraphics[height = 2.7cm]{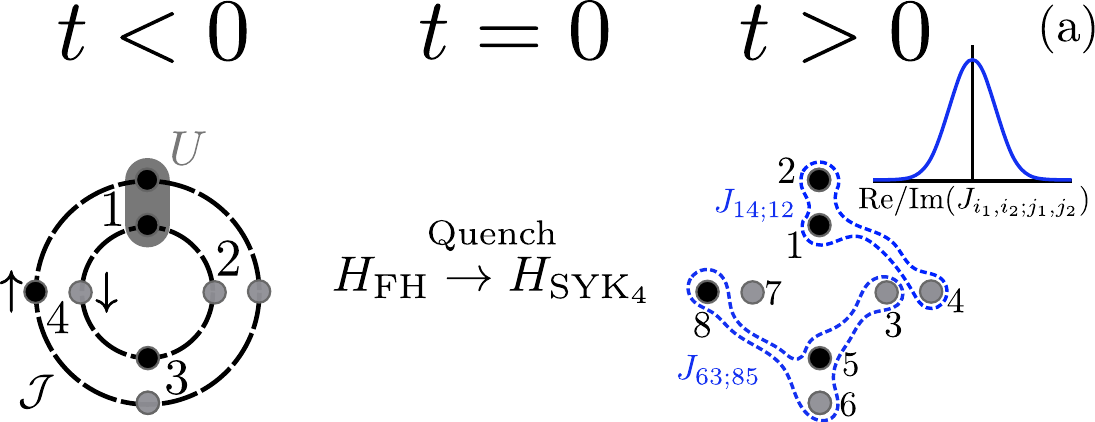}
 \vspace{0.3cm}
 \includegraphics[height = 5.7cm]{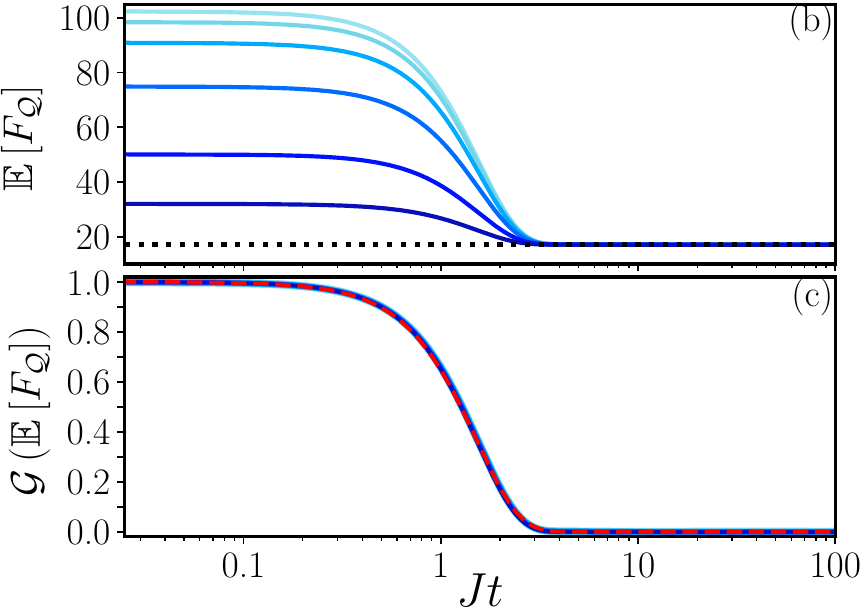}
 \caption{
	   Universal super-exponential equilibration dynamics of the QFI, 
	   $F_{\mathcal{Q}}$, under the complex SYK$_4$ Hamiltonian.  
	   (a) Illustration of the quench protocol.  
	   Left: Initial states are chosen as ground states of the Fermi--Hubbard (FH) 
	   Hamiltonian for different values of $U/\mathcal{J}$ (other generic initial 
	   states yield equivalent results). 
	   The black and gray circles respectively represent occupied and empty 
	   fermionic modes. 
	   Dashed lines illustrate hopping of fermions between nearest-neighbor 
	   sites. 
	   Right: The system is evolved under the SYK$_4$ Hamiltonian, where spinless 
	   fermions (black circles) can hop to any empty 
	   fermionic mode (light gray circles). The disordered
	   interaction strengths $\left\{J_{i_{1}i_{2};j_{1}j_{2}}\right\}$ 
	   are randomly sampled from independent Gaussian distributions.  
	   (b) QFI averaged over $400$ disorder realizations, 
	   $\mathbb{E}\left[F_{\mathcal{Q}}\right]$, 
	   computed with respect to the operator $\hat{R}$ defined in 
	   Eq.~(\ref{e:stagmagSYK}). Initial states from darker to lighter shade of blue 
	   are for $U/\mathcal{J} = 0, 2, 4, 6, 8$, and $10$. 
	   The system equilibrates fast to the expectation value of the 
	   Gibbs infinite temperature state (dotted black line).
	   (c) Universality in the dynamics is revealed by rescaling to  
	   $\mathcal{G}\left(\mathbb{E}\left[F_{\mathcal{Q}}\right]\right)$, 
	   as given in Eq.~(\ref{e:rescaling}). 
	   Very good agreement is found to a Gaussian fit, 
	   $\mathrm{exp}\left[-(Jt/\tau)^2\right]$, with a fast decay constant of 
	   $\tau = 1.52$ (dashed red curve). 
	   Data for $Q=8$ fermions occupying $N = 16$ fermionic modes. 
	   }
 \label{f:fig_schem_univ_fq}
\end{figure}

\section{Quench protocol}\label{s:quench_protocol}

We are interested in the disorder-averaged out-of-equilibrium dynamics 
generated
by the SYK$_{q}$ model for spinless complex fermions, following 
the quench protocol sketched in Fig.~\ref{f:fig_schem_univ_fq}a. The family of 
SYK$_{q}$ models is characterized by $(q/2)$-body random all-to-all interactions, 
where $q/2\in Z^{+}$. 
In this paper, we present in detail the dynamics of the SYK$_{4}$ model, but we stress 
that the salient features of this equilibration process are qualitatively generic to 
other values of $q$~\cite{bandyopadhyay_2021}.
The Hamiltonian of an instance of the ensemble reads (see 
App.~\ref{app:DetailsOnSYKModel} for details)~\cite{gu_2020,baldwin_2020}
\begin{equation}
\hat{H}_{\mathrm{SYK}_{4}} = \frac{1}{(2N)^{\frac{3}{2}}}
  \sum_{\substack{i_1,i_2,j_1,j_{2} = 1}}^{N} J_{i_{1}i_{2};j_{1}j_{2}} 
  \hat{c}^{\dagger}_{i_{1}}\hat{c}^{\dagger}_{i_{2}} 
  \hat{c}_{j_{1}}\hat{c}_{j_{2}}. 
\label{e:Hsyk4}
\end{equation}
There are $N$ spinless fermionic modes with creation, annihilation, and 
occupation number operators that satisfy canonical anticommutation relations 
and are denoted, respectively, by $\hat{c}_{i}^{\dagger}$, $\hat{c}_{i}$, 
and $\hat{n}_{i}$. 
The interaction strengths $\{J_{i_{1}i_{2};j_{1}j_{2}}\}$ are complex Gaussian random 
variables, and we denote the disorder average over realizations by 
$\mathbb{E}[...]$.

As initial states $\ket{\psi(0)}$, prepared at time ${t<0}$, we consider for convenience 
ground states of the one-dimensional spinful Fermi--Hubbard (FH) 
model~\cite{hubbard_1963, fradkin_2013} given by the Hamiltonian  
$ \hat{H}_\mathrm{FH} = -\mathcal{J}\sum_{\ell=1}^{N/2} \sum_{\sigma 
    =\uparrow,\downarrow} \left( \hat{c}_{\ell,\sigma}^\dagger 
\hat{c}_{\ell+1,\sigma}  + \mathrm{H.c.}\right) 
+ U \sum_{\ell=1}^{N/2}  \hat{n}_{\ell,\uparrow}\hat{n}_{\ell,\downarrow}$.
In the FH model, physical modes are given by $N/2$ spatial lattice sites, 
$\ell=1, \dots, N/2$, and additional spin degrees of freedom (represented by 
the arrows) not 
present in the SYK model. 
As the family of SYK$_q$ models consists of zero-dimensional models (due to 
site-independent all-to-all interactions), the mapping of FH modes to the SYK 
modes is arbitrary. We choose 
here 
$\{\ell,\uparrow\}\leftrightarrow i=2\ell$ and 
$\{\ell,\downarrow\}\leftrightarrow i=2\ell-1$.
The FH ground states are chosen as a representative case of initial states 
    whose operator expectation values differ significantly from the steady-state
values.

In our numerics, we employ various choices of the boundary condition,
\footnote{We performed computations for periodic, anti-periodic, and open 
boundary conditions. In this article, we present the dynamics of initial states 
obtained with periodic and anti-periodic boundary conditions when $N/4$ is odd 
and even, respectively.} 
ratio between onsite interaction and hopping strengths $U/\mathcal{J}$, total fermion 
number $1 \leq Q \leq N$, and total magnetization (we focus mostly on the case of 
half-filling, $Q=N/2$, and zero magnetization, where there are $N/4$ fermions in each 
spin sector).
We emphasize that the above choice of the initial Hamiltonian 
is only for convenience of preparing initial states that cover a range of 
parameters in a strongly-correlated system.
Choosing other generic initial states does not modify our findings.

Once the system is prepared in the initial state $\ket{\psi(0)}$, we perform a global 
quench at $t=0$ to the SYK$_{4}$ model and track the state's subsequent unitary time 
evolution (here and throughout we set $\hbar=1$)
\begin{equation}
\ket{\psi(t)}=e^{-i\hat{H}_{\mathrm{SYK}_{4}}t}\ket{\psi(0)}. 
\label{e:unievo}
\end{equation}
We average the time series over multiple disorder realizations in order to filter out the 
salient, realization-independent features of the equilibration dynamics. 
To test the generality of our findings, we study a variety of observables, in 
particular the QFI (see next section) and higher-order correlators (see 
Sec.~\ref{s:uni_dynSYK4} and App.~\ref{app:higher_moments}) corresponding to 
the staggered magnetization, as well as other few-body ($4$-local) operators 
and a non-diagonal generator for the QFI (see 
App.~\ref{app:other_observables}).

As a main result of our study, the disorder-averaged evolution of the 
considered few-body observables shows universality under the following rescaling  
\begin{equation} 
 \mathcal{G}\left(f(t)\right) 
	= \frac{f(t)-\overline{f(t)}}{f(0)-\overline{f(t)}},
\label{e:rescaling}
\end{equation}
where $\overline{f(t)}$ represents the long-time average of the function $f(t)$ 
computed over a time window starting at $t_0$ and with duration $T$, i.e.,
\begin{equation}
 \overline{f(t)} = \frac{1}{T}\int_{t_0}^{t_0 +T} f(t) dt .
 \label{e:longtimeav}
\end{equation}
Unless explicitly stated, we consider $Jt_0 = 50$ and $J(t_0 + T) = 100$ for the exact 
diagonalization results presented in this paper. 
It is to be noted that the SYK model is only parameterized by the variance of 
the random all-to-all interaction strengths [see Eq.~\eqref{e:syk_statistics}].
Therefore, the universality in the evolution of observables can be probed as 
the insensitivity to the initial conditions, which becomes evident under the 
rescaling in Eq.~\eqref{e:rescaling}, as is discussed throughout this 
manuscript.

\section{Universal super-exponential equilibration dynamics}
\label{s:unidynqfi}

To illustrate the universal equilibration dynamics, we present 
here results for the QFI evolved under the SYK$_{4}$ Hamiltonian.
The QFI is an observable of central relevance in quantum 
sensing~\cite{PezzeSmerzi_2014,review_Degen_etal17,review_Pezze_etal18}, which 
can witness multipartite 
entanglement in quantum many-body systems at zero and finite
temperatures~\cite{Toth_2012,Hyllus_etal12, Hauke_etal16, 
gabbrielli_2018, AlmeidaHauke20}.
Interestingly, like the out-of-time-order correlators~\cite{foini_2019, 
chan_2019}, this quantum information theoretic measure can distinguish a pure 
eigenstate of an ETH-obeying Hamiltonian from the corresponding Gibbs thermal 
state~\cite{brenes_2020}. 

In the present context of pure states, the QFI with respect to an observable 
$\hat{O}$ is simply proportional to its variance
\begin{equation}
 F_{\mathcal{Q}}[\hat{O}](t) 
  = 4\left(\bra{\psi(t)}\hat{O}^{2}\ket{\psi(t)}
	  -\bra{\psi(t)}\hat{O}\ket{\psi(t)}^2\right). 
 \label{e:fishinfo}	  
\end{equation}
In this section, we consider the staggered magnetization, which in the FH model is defined as $\hat{O}_{\rm SM} = \sum_{\ell=1}^{N/2} (-1)^{\ell}(\hat{n}_{\ell\downarrow} -\hat{n}_{\ell\uparrow})$, and which in the SYK model translates to 
\begin{equation}
\hat{R} = \sum_{i=1}^{N/2} (-1)^{i}\hat{\kappa}_{i} =
\sum_{i=1}^{N/2} (-1)^{i}(\hat{n}_{2i-1}-\hat{n}_{2i}). 
\label{e:stagmagSYK}
\end{equation}
The $\hat{\kappa}_{i}$ denote $2$-local operators which we use to construct the 
$4$-local operators discussed in App.~\ref{app:other_observables}.

The time evolution of the disorder-averaged QFI, $\mathbb{E}\left[F_{\mathcal 
Q}\right]$, 
computed with respect to the operator $\hat{R}$ is shown in 
Fig.~\ref{f:fig_schem_univ_fq}b. 
The considered initial states are the symmetry unbroken ground states of the FH model for
$U/\mathcal{J}= 0, 2, 4, 6, 8,$ and $10$, respectively (dark to light shading). These 
values include a non-interacting initial system for $U/\mathcal{J} = 0$, as well as 
strongly interacting systems at larger values of $U/\mathcal{J}$. As a result, the initial
states are characterized by a varying amount of multipartite entanglement that is 
witnessed by the QFI.\footnote{We note that the scaling of the QFI with system size $N$ depends on the amount of multipartite entanglement of the state \cite{review_Pezze_etal18}: For separable states
     states, $F_{\mathcal{Q}} \leq N$, whilst for genuinely $N$-partite entangled states $F_{\mathcal{Q}} \leq N^2$.}
At short times, the system still retains memory of the initial 
state. However, the completely disordered all-to-all interactions of the SYK model lead 
to a quick loss of this memory, and already at a time of about  $Jt\approx 4$ the QFI 
equilibrates to a steady state value that is independent of the initial state. This rapid 
equilibration is reminiscent of the fast scrambling characteristic of the 
model, and also bears similarities to the relaxation curves derived in 
Ref.~\cite{Reimann_2016} for the out-of-equilibrium dynamics of isolated 
quantum  many-body systems. There it is shown that rapid, non-exponential  
equilibration is expected for, in a random matrix sense, typical Hamiltonians 
and observables.

The attained steady state value matches with the one obtained from the infinite 
temperature Gibbs state (horizontal dashed black line in 
Fig.~\ref{f:fig_schem_univ_fq}b)
\begin{equation}
 \hat{\rho}_{\infty} = \left.\frac{{\rm e}^{-\beta\hat{H}}}
	               {Z}\right\rvert_{\beta = 0}
		     = \frac{\mathds{1}}{D},
 \label{e:infin_temp}		     
\end{equation}
where $\mathds{1}$ is the identity operator, $Z$ is the partition 
function and $D$ is the Hilbert space dimension 
determined by $N$ and $Q$. 
This finding suggests that the overlaps between a generic initial state 
and energy eigenstates of the SYK$_4$ Hamiltonian are uniformly distributed over the spectrum. This is substantiated by computing the Kullback--Leibler 
divergence, $D_\mathrm{KL}(P(E) \Vert Q(E))$, between the uniform distribution 
$Q(E)=1/D$ and the initial states' amplitude distribution     
$P(E)=\abs{ \langle \psi(0) \vert E \rangle }^2$ with respect to the     
energy basis $\{\ket{E}\}$.\footnote{For $N = 8$ and $N = 12$ systems, 
$\eave{D_\mathrm{KL}} =  0.0997 \pm 0.0154$ and $0.0627 \pm 0.0028$, respectively.
This indicates that the initial states are almost uniformly distributed, and the 
uniformity of $P(E)$ improves with increasing $N$. The quoted values are for the FH 
initial state $U/\mathcal{J} = 10$, and are representative of all considered 
initial states.}

We note that even though the steady state $ \hat{\rho}_{\infty}$ has vanishing QFI (see for instance Ref.~\cite{PezzeSmerzi_2014}), it
is nevertheless an interesting question of how the system reaches that point, starting from initial
states with different amounts of quantum correlations. Indeed, we find that this equilibration dynamics is universal within numerical precision, 
as one can expose by rescaling
$\mathbb{E}\left[F_{\mathcal Q}\right]$ according to Eq.~(\ref{e:rescaling}). The 
dynamics of the rescaled curves are shown in Fig.~\ref{f:fig_schem_univ_fq}c. 
All the curves collapse throughout the dynamics, independent of the initial state. 
In addition, the universal curve can be well approximated by a Gaussian (red 
dashed curve 
in Fig.~\ref{f:fig_schem_univ_fq}c), 
with a fast decay constant of $\tau = 1.52$. Thus, 
under the Hamiltonian evolution of the SYK$_{4}$ model, the disorder-averaged 
QFI exhibits 
universal and super-exponential equilibration dynamics. 
%

    We note that within the context of random matrix theory it is well known that a Gaussian temporal evolution can occur in the survival probability.
    For instance, this happens for quantum quench dynamics under Wigner random banded matrices and two-body random ensembles  (see for instance Ref.~\cite{FlambaumIzrailev_2001}, and Review~\cite{Borgonovi_etal2016} and references therein), where the latter is a specific case of embedded random matrix ensembles~\cite{vyas_2017}.
    The same behavior has also been studied for a generic disordered interacting spin model~\cite{tavora_2016}.
    In these cases, the survival probability initially decays as a Gaussian, followed 
    by a regime in which it shows oscillations with a power-law 
    envelope~\cite{tavora_2016}. 
    The oscillatory behavior is also seen in the evolution of the spectral form factor of SYK 
    models~\cite{cotler_2017, sonner_2017}.
    Our numerics illustrate the above features for the survival probability  
    (see Fig.~\ref{f:fidelity_log_log}). 
    Indeed, for $q=4$ the SYK model can be interpreted as a two-body random ensemble (albeit without a mean-field contribution), and more generally as an embedded random matrix ensemble for $q\geq 2$ [see Eq.~\eqref{e:Hsykq}], and as such the above features are expected.
    In contrast to this many-body observable, for the few-body observables considered in this study, we find the Gaussian decay followed by a 
    marginal domain in which an indication of a power-law tail is 
    obtained, but which appears to diminish with system size (see Fig.~\ref{f:fig_moment2_log_log}).

\section{Dissipative ensemble dynamics}\label{s:master_eq}

In this section, we present the key results of the open-system formalism, which we use to 
understand the main features observed in the average dynamics of the unitary ensemble 
discussed in the previous section. Due to its generality for treating disorder ensemble 
averages, this formalism is of interest also independent of the application to 
the present scenario. 
The derivation presented here is formally written for time-dependent random 
processes. 
We stress, however, that we allow for time-dependent stochastic processes in Eq.~\eqref{e:Hgeneric} only to make our formalism applicable to more general Hamiltonians.
The resulting evolution equations apply also in the limit of static processes, such as those defining the SYK model. 
The approach presented here has the advantage of treating quenched disorder and temporally fluctuating 
noise on equal footing, enabling an application to a large variety of settings. 
The interested reader may find further details and the explicit derivation for static 
processes in App.~\ref{app:MEapp}.

Our aim is to directly study the dynamics of the ensemble's density matrix 
$\tilde{\rho}(t) \equiv \eave{\hat{\rho}(t)}$ via the ensemble averaged von 
Neumann equation (EAVNE), where the time-evolution of each state 
$\hat{\rho}(t)$ is generated by a realization of the general closed system 
Hamiltonian
\begin{equation}\label{e:Hgeneric}
\hat{H}(t) = \hat{H}_0(t) + \sum_\alpha \hat{H}_\alpha(t) .
\end{equation}
Here, $\hat{H}_0(t)$ is a disorder-free contribution, which in general can be time 
dependent, whereas the terms
\begin{equation}
\label{e:H_alpha_generic}
\hat{H}_\alpha(t) = \sum_{l_\alpha} \xi_{l_\alpha}^{(\alpha)}(t) 
\hat{h}^{(\alpha)}_{l_\alpha}, 
\end{equation}
capture the dynamics due to disorder or noise. The index $\alpha$ is used 
to distinguish different subsets of Hermitian operators 
$\hat{h}^{(\alpha)}_{l_\alpha}$, and the operators within a subset are labeled 
by the (multi-)index $l_\alpha$.\footnote{Whilst the distinction via index 
$\alpha$ is not strictly necessary for our 
derivation, it does facilitate translation of our general results to specific models in 
which such a distinction may naturally arise. For example, in a system of spins 
arranged on a lattice, $\alpha=1$ could refer to a disordered external potential and 
$\alpha=2$ to a noisy drive. For either, the operator label $l_\alpha$ 
would refer to the site index of the spins.} 
In particular, upon rewriting the SYK$_{4}$ Hamiltonian in the generic 
form of Eq.~\eqref{e:H_alpha_generic}, we identify three operator subsets, as shown 
in Sec.~\ref{s:subsec_syk4}. We assume the functions $\xi_{l_\alpha}^{(\alpha)}(t)$ to 
describe a Gaussian process 
possessing---without loss of generality---vanishing cross-correlations, 
$\eave{\xi_{l_\alpha}^{(\alpha)}(t) \xi_{l_\beta}^{(\beta)}(t') } = 0$ for  
$\alpha \neq \beta$, so that their correlation tensor is given by 
$F^{(\alpha)}_{l_\alpha,k_\alpha}(t,t') 
\equiv \eave{\xi_{l_\alpha}^{(\alpha)}(t) \xi_{k_\alpha}^{(\alpha)}(t')}$. 
Formally, the SYK$_{4}$ model defined in Eq.~\eqref{e:Hsyk4} corresponds to setting 
$\hat{H}_0(t)=0$ and taking all 
$\xi_{l_\alpha}^{(\alpha)}(t)$ to be time independent, in which case
$F^{(\alpha)}_{l_\alpha,k_\alpha}(t,t')$ is constant with respect to time. 
To keep the formalism general, we will specialize to this case only further below.

Consider the EAVNE generated by averaging over multiple realizations of the 
Hamiltonian in Eq.~\eqref{e:Hgeneric}, 
\begin{equation}\label{e:vNensave}
\partial_t \tilde{\rho}(t) 
= -i  \comm{\hat{H}_0(t)}{ \tilde{\rho}(t) } 
-i\sum_{\alpha ,l_\alpha} \comm{ \hat{h}^{(\alpha)}_{l_\alpha}  }{ 
    \eave{\xi_{l_\alpha}^{(\alpha)}(t) \hat{\rho}(t)}  } .
\end{equation} 
To proceed, one needs to handle the correlations 
$\eave{\xi_{l_\alpha}^{(\alpha)}(t) \hat{\rho}(t)} $.
These are non-trivial since the density matrix is---by virtue of the 
von~Neumann equation---a functional $\hat{\rho}[\xi,t]$ of the Gaussian 
processes
$\xi_{l_\alpha}^{(\alpha)}(t)$.
The simplicity of our framework rests upon use of the Novikov--Furutsu theorem 
\cite{Novikov65,Furutsu63,Furutsu72,KlyatskinTatarskii72}, which provides an 
exact expression of these correlations in terms of  
$F^{(\alpha)}_{l_\alpha,k_\alpha}(t,t')$ as 
    \begin{equation}\label{e:novikov}
    \eave{\xi^{(\alpha)}_{l_\alpha}(t) \hat{\rho}[\xi,t]} = \sum_{k_\alpha} 
    \int_0^\infty dt' 
    F^{(\alpha)}_{l_\alpha,k_\alpha}(t,t') 
    \eave{\frac{\var \hat{\rho}[\xi,t]}{\var \xi^{(\alpha)}_{k_\alpha}(t')}} .
    \end{equation}
An explicit expression for the functional derivative can be obtained from the 
integrated von~Neumann equation. 
Formally, this yields an infinite series in which the $n$th term ($n \geq 1$) contains $n-1$ time integrals over $n$ nested commutators, from which the exact functional derivative can be obtained in principle.
For a systematic study of the role of the higher order terms, we refer the reader to our follow-up work Ref.~\cite{Paviglianiti_etal2022}.
Here, we retain only the lowest order contribution ($n=1$), which reads 
    \begin{equation}\label{e:funcderiv}
    \frac{\var \hat{\rho}[\xi,t]}{\var \xi^{(\alpha)}_{k_\alpha}(t') }  \simeq 
    -i\comm{\hat{h}^{(\alpha)}_{k_\alpha}}{\hat{\rho}[\xi,t']} \Theta(t-t'),
    \end{equation}
where the step-function $\Theta$ arises from causality.    
Substituting Eqs.~\eqref{e:novikov} and \eqref{e:funcderiv} into 
Eq.~\eqref{e:vNensave}, we obtain 
the evolution equation
    \begin{equation}\label{e:preMarkov}
    \begin{split}
    \partial_t \tilde{\rho}(t)\! =&\! -i 
    \comm{\hat{H}_0(t)}{\tilde{\rho}(t)}\!\\
    &-\! \sum_{\alpha,l_\alpha, 
        k_\alpha} 
        \comm{\hat{h}^{(\alpha)}_{l_\alpha}}{\comm{\hat{h}_{k_\alpha}^{(\alpha)}
         } 
        {\int_{0}^{t}dt'F^{(\alpha)}_{l_\alpha,k_\alpha}(t,t') 
        \tilde{\rho}(t')}} .
    \end{split}
    \end{equation}

\begin{figure}[ht!]
    \includegraphics[width = 
    \linewidth]{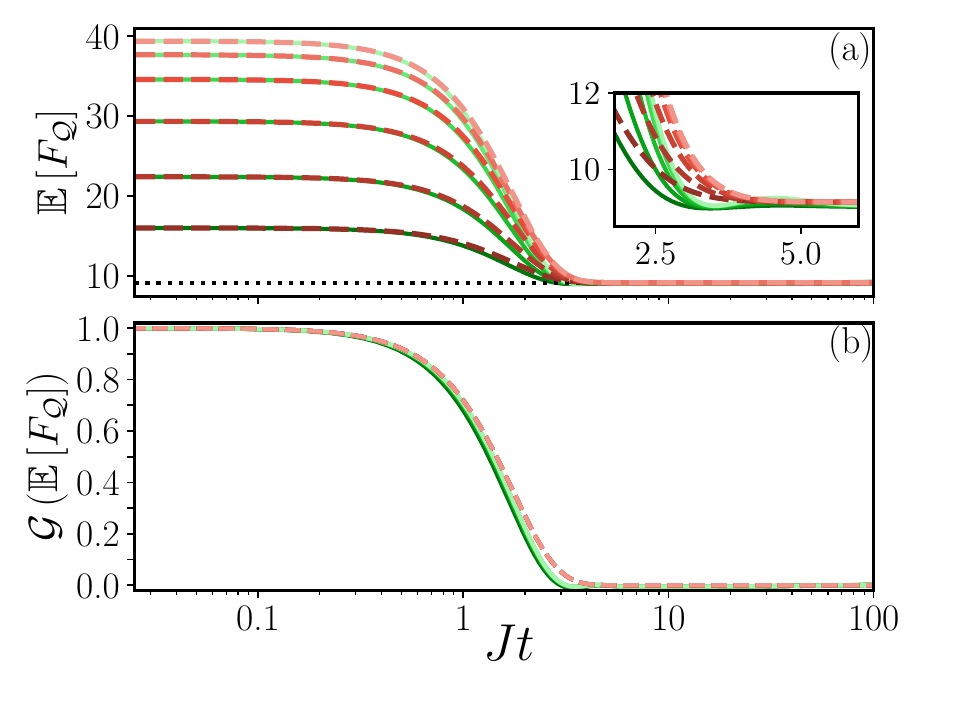}
    \caption{
    Comparison of ED and ME results for the QFI of the staggered 
    magnetization $\hat{R}$ in the SYK$_4$ model for $N=8$. ED curves are 
    averaged over $90000$ disorder realizations. 
    For both ME (red, dashed) and ED (green, solid) curves, dark to light 
    shading corresponds to the different initial states of  
    Fig.~\ref{f:fig_schem_univ_fq}.
    The black (dotted) 
    line shows the analytically predicted steady-state value of 
    Eq.~\eqref{e:ssvalueME} for the half-filling sector.
    (a) For each initial state, the ME simulation reproduces the 
    dynamics of the exact numerics very well.
    There is a discrepancy at intermediate times due to non-Markovian effects 
    and higher-order correlations, not captured by the approximate ME (inset).
    (b) The ME reproduces the collapse to a universal curve under 
    the rescaling defined in Eq.~\eqref{e:rescaling}, without any free fit 
    parameter.
    The slight spread in the rescaled ED curves is due to statistical 
    fluctuations, which for the QFI are suppressed for larger system 
    sizes, as 
    can be seen from a comparison with Fig.~\ref{f:fig_schem_univ_fq}c, which 
    shows QFI dynamics computed via ED for $N=16$. 
    }
    \label{f:qfi_meVSed}
\end{figure}

This evolution equation is not exact, due to our use of the approximate 
functional derivative given by Eq.~\eqref{e:funcderiv}. 
Such a leading order truncation amounts to the well-known decorrelation 
assumption\footnote{This is readily seen within the interaction 
picture generated by $\hat{H}_0(t)$, where the truncated functional derivative 
now acts on the transformed functional given by the \emph{interaction} picture 
density matrix.}
typically made in the analysis of stochastic evolution 
equations \cite{GardinerZoller_textbook,vanKampen_textbook}.
The remaining time integral in Eq.~\eqref{e:preMarkov} is known as a Bourret 
integral \cite{Bourret1962, Dubkov1977}. 
While the decorrelation assumption becomes exact in the limit of white noise, 
for non-Markovian noise it corresponds to an expansion controlled by the noise 
correlation time \cite{vanKampen1974}. 
One may then wonder what justifies this approximation 
(see also Fig.~\ref{f:qfi_meVSed}) for the present disordered system, which has 
an infinite correlation time. 
The reason can be attributed to the chaoticity of the SYK$_{4}$ model, rigorous 
proof of which remains an open question for future investigations.
Each term of the Hamiltonian in Eq.~\eqref{e:H_alpha_generic} can be thought of 
as an independent noise process governing the evolution of the density 
operator. 
Then, in the presence of a \emph{large} number of such processes---as in the 
SYK$_4$ model---one may expect the correlations between the density operator 
and any individual process to be 
strongly suppressed.
Viewed differently, the decorrelation assumption can be seen as a 
linear-response approximation \cite{Visscher1974,Schekochihin01}, i.e., the 
response of the state $\hat{\rho}$ at time $t$ towards a perturbation with 
$\xi^{(\alpha)}_{l_\alpha} \hat{h}^{(\alpha)}_{l_\alpha}$ at an earlier time 
$t_1$ is taken into account only to linear order.
In the context of Kubo's celebrated linear response theory \cite{Kubo57}, it is 
well known that averaging effects due to chaos lead to a superb success much 
beyond the regime of applicability predicted by na\"{i}ve estimates 
\cite{Visscher1974,vanVelsen1977,Kubo1985,vanVliet1988,Goderis1991}. 
In the present context, the observed success of the linear approximation can 
be seen as a manifestation of the strong effects of quantum chaos in the 
SYK$_{4}$ model.

The master equation as given by Eq.~\eqref{e:preMarkov} is still rather 
unwieldy, since it is not local in time.
We thus perform a Markov approximation leading us to a Lindblad master equation 
in non-diagonal form \cite{BreuerPetruccioneTextBook} governed by a 
time-dependent Liouvillian superoperator 
\begin{equation}\label{e:me}
\mathcal{L}(t)\tilde{\rho}(t) = -i \comm{\hat{H}_0(t)}{ \tilde{\rho}(t) } + 
\sum_{\alpha} 
\mathcal{D}^{(\alpha)}(t) \tilde{\rho}(t),
\end{equation}
with Hermitian dissipator
\begin{equation}\label{e:dissip}
\begin{split}
\mathcal{D}^{(\alpha)}(t) \tilde{\rho}(t) =& \sum_{l_{\alpha}, k_{\alpha}} 
2 f^{(\alpha)}_{l_\alpha, 
    k_\alpha}(t) \\
&\times \left( \hat{h}^{(\alpha)}_{l_\alpha} \tilde{\rho}(t)
\hat{h}^{(\alpha)}_{k_\alpha} - \frac{1}{2} 
\acomm{\hat{h}^{(\alpha)}_{k_\alpha} 
    \hat{h}^{(\alpha)}_{l_\alpha}}{ \tilde{\rho}(t) }  \right) ,
\end{split}
\end{equation}
in which we have defined     
\begin{equation}\label{e:Fint}
f^{(\alpha)}_{l_\alpha, k_\alpha}(t) = 
\int_{0}^{t}dt' F^{(\alpha)}_{l_\alpha,k_\alpha}(t,t') .
\end{equation}

Equations~\eqref{e:me}--\eqref{e:Fint} form the final evolution equations
of this section. They are valid under a Bourret--Markov approximation for the 
generic Hamiltonians of Eq.~\eqref{e:Hgeneric} with disorder and/or noise 
contributions.
We reiterate that these effective evolution equations do not require the presence of noise (see App.~\ref{app:MEapp}, fourth paragraph), and that the master equation is a result of averaging over an ensemble of disorder realizations.
Whilst each individual disorder realization evolves unitarily, the ensemble 
evolves like an open system, with a dynamics that is approximately generated 
by the Liouvillian $\mathcal{L}(t)$.
The coherent and dissipative processes that constitute $\mathcal{L}(t)$ can be 
read-off immediately from the system's Hamiltonian. 
The corresponding dissipation rates are entirely determined by the disorder 
statistics $F^{(\alpha)}_{l_\alpha,k_\alpha}(t,t')$ via Eq.~\eqref{e:Fint}.
Similarly, while each realization preserves the purity of the initial state, 
the Hermitian jump operators of the master equation generate ensemble dynamics 
that are purity decreasing~\cite{Manzano20,Lidar_etal2006}, and thus drive the 
ensemble from a pure to a mixed state. In particular, for the SYK$_{4}$ model 
with large enough $N$, the ensemble equilibrates to the 
infinite-temperature state $\hat{\rho}_{\infty}$ given in 
Eq.~\eqref{e:infin_temp} \cite{Kraus08,Lidar_etal2006}.
Note, however, that whilst the Hermitian jump 
operators of Eq.~\eqref{e:me}~and~\eqref{e:dissip} ensure that 
$\hat{\rho}_{\infty}$ is a steady state of the Liouvillian
dynamics, for an arbitrary Hamiltonian as given by 
Eq.~\eqref{e:Hgeneric}, the steady-state of $\mathcal{L}(t)$ need not be unique 
in general \cite{Minganti18,Tindall19}.

Besides clarifying the nature of the steady-state, the Liouvillian dynamics also 
reproduce the rapid, super-exponential equilibration of the QFI observed in the SYK$_4$ 
quench dynamics of Sec.~\ref{s:quench_protocol}: 
In the above equations, the case of static disorder is captured by a ``noise'' 
correlation that is constant in time, so that $ 2 f^{(\alpha)}_{l_\alpha, 
k_\alpha}(t) 
= 2t \eave{ \xi_{l_\alpha}^{(\alpha)} \xi_{k_\alpha}^{(\alpha)} } $. 
Thus, under the Bourret--Markov approximation the SYK$_4$ model (or indeed any SYK$_q$ 
model) is governed by dissipation rates that grow linearly in time. 
Consequently, one can factor the Liouvillian as $\mathcal{L}(t) = 2t \mathcal{D}$, which 
naturally yields the super-exponential time-evolution 
\begin{equation}\label{e:formal_time_evo}
\tilde{\rho}(t)=\mathcal{T} e^{\int_{0}^{t} dt' 2t' \mathcal{D} 
}\hat{\rho}(0)=e^{t^2 
    \mathcal{D} }\hat{\rho}(0) ,
\end{equation}
where $\mathcal{T}$ denotes time-ordering.

Figure~\ref{f:qfi_meVSed}a shows simulations of the QFI evolution generated by 
the above master equation.
In general, the ME as developed here may not be used to study the ensemble 
average of the QFI, due to the term $\bra{\psi(t)}\hat{O}\ket{\psi(t)}^2$ [see 
Eq.~\eqref{e:fishinfo}] that is non-linear in the density matrix. However, 
for the initial zero magnetization states considered here, exact numerics show 
that the first moment fluctuates around zero, such that in this case one may use the 
ME to approximate the dynamics of the QFI. Indeed, the agreement with the ensemble 
averaged ED results is striking, and we emphasize that no fit parameter has been 
used (nor is one available in the above formalism) to 
achieve this agreement.
The Liouvillian dynamics also 
reproduce the universality of the QFI with 
respect to different initial states, as shown in Fig.~\ref{f:qfi_meVSed}b. 
A discrepancy at 
intermediate times can be attributed to the decorrelation and Markov 
approximations of the master equation. These approximations can be expected to hold especially 
at early times, as can be seen from a short-time expansion, and at late times 
when the system enters a steady state (a unique steady state is the one which is reached 
independent of the precise trajectory, so correlations to the exact noise process and 
memory about previous times can be expected to be negligible).
Indeed, the ME  provides excellent 
agreement at early and late times, and successfully captures the overall trend 
of the dynamics even at intermediate times.

Even more, the Liouvillian formalism allows us 
to study the origin of the universal dynamics. In Sec.~\ref{s:MEuniversality}, 
we analyze how different states and observables decompose over the eigenspaces of 
$\mathcal{L}(t)$, showing how these distributions conspire in the 
presented scenario to select a single 
time-scale, universal across different initial states. 

As a final remark, the fact that disorder induces dephasing 
between members of an ensemble---and thus leads to effective open-system 
evolution equations even in the absence of a heat bath---is long known; in 
particular for special single-body cases such as 
classical disordered dipoles and harmonic oscillators \cite{vanKampen1974} or 
single 
atoms coupled to a photon field~\cite{ghoshal_2020}. 
Here, our aim was the derivation of a general framework for quantum many-body systems. 
Such a 
platform for the evolution 
of disorder-averaged density operators has been rigorously developed previously 
\cite{Kropf_etal16}, based on a matrix formalism \cite{andersson_2007,hall_2014}. 
Our derivation based on the Novikov--Furutsu theorem is simpler but nevertheless general, 
as the assumptions we made are not fundamental, but rather of practical nature: 
The Novikov--Furutsu formalism can be extended to non-Gaussian stochastic 
processes \cite{KlyatskinTatarskii72, haenggi_1978}, the decorrelation 
assumption may be 
lifted in favor of an infinite series of terms in Eq.~\eqref{e:funcderiv}
\cite{vanKampen1974,Schekochihin01}, and 
the Markov approximation is---at least on a formal level---not necessary in the 
derivation of the evolution equations.
The present framework has the additional feature that disorder and noise 
processes can be 
treated on equal footing, within the same master equation, without any further 
complications of the formalism. 
That property enables us to tap into the vast literature employing the 
Novikov--Furutsu theorem in the context of noise with finite correlation time 
(see, e.g., Refs.~\cite{Montiel_etal19, Chenu_etal2017, budini_2000, 
budini_2001}). 

To summarize this section, the Novikov--Furutsu theorem enables us to derive 
a master equation, for the ensemble average, that provides a general framework 
for disorder-averaged quantum many-body systems.
For the case of the SYK$_{4}$ model, it yields a series of analytic insights 
into the out-of-equilibrium dynamics, such as the steady state reached at late 
times, the approximately Gaussian decay, and the universal behavior across 
different initial states.

\begin{figure}[ht!]
 \includegraphics[height = 6.25cm]{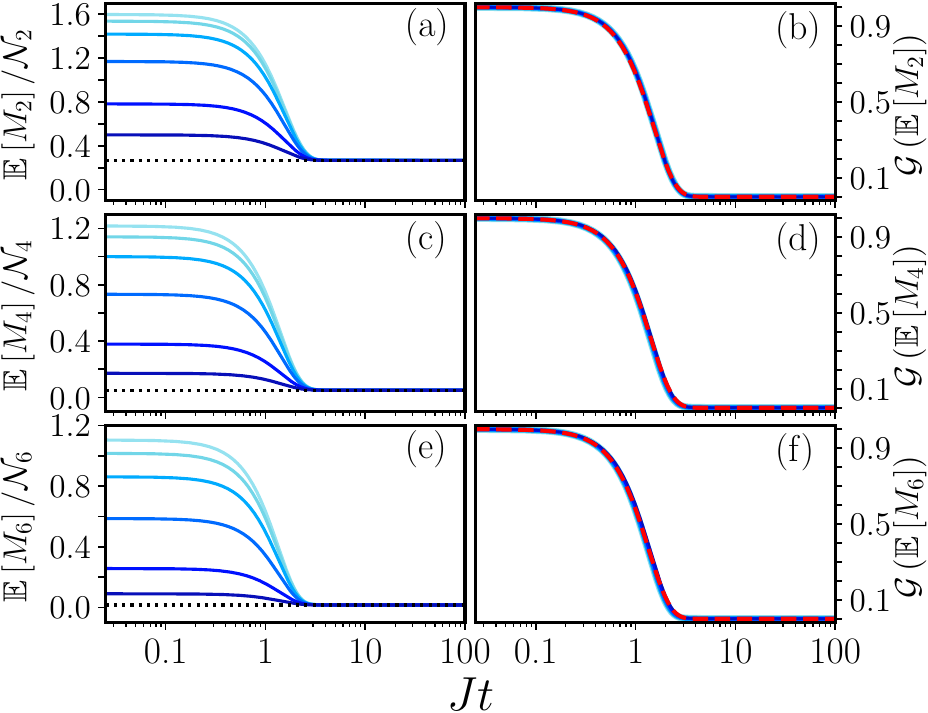}
 \caption{
	   Disorder-averaged universal equilibration dynamics of the 
	   $k$th moment, $M_{k}$, of the operator $\hat{R}$ under the SYK$_4$ 
	   Hamiltonian. Data is for $Q=8$ fermions occupying $N = 16$ modes.  
	   Left column: Dynamics of $M_{2}$, $M_{4}$, and $M_{6}$ averaged over 
	   $400$ disorder realizations, with only a simple normalization 
	   $\mathcal{N}_{k}$ as in 
	   Eq.~(\ref{e:scale_fac}) for visualization. 
	   Right column: Corresponding dynamics rescaled by the function 
	   $\mathcal{G}$ given in Eq.~(\ref{e:rescaling}). 
	   Curves for different initial states (chosen as in 
	   Fig.~\ref{f:fig_schem_univ_fq}) collapse. 
	   The dotted black lines in (a), (c), and (e) 
	   mark the values of the operator moments calculated with respect to 
	   the Gibbs infinite temperature state. The dashed red curves in (b), 
	   (d), and (f) correspond to Gaussian fits, 
	   $\mathrm{exp}\left[-(Jt/\tau)^2\right]$, with $\tau = 1.52, 1.42, 
	   1.35$, respectively. Similar results for $k= 8, 10,$ and $12$ can be 
	   found in Fig.~\ref{f:fig_moments_higher}.
          }
 \label{f:fig_moments}
\end{figure}

\section{Universal evolution of operator moments under SYK$_{4}$}
\label{s:uni_dynSYK4}

To corroborate the generality of the above findings, in this section we 
consider the $k$th moment, $M_{k}(t) = \bra{\psi(t)}\hat{O}^{k}\ket{\psi(t)}$, 
of the operator $\hat{O}=\hat{R}$ defined in Eq.~\eqref{e:stagmagSYK} 
(in App.~\ref{app:other_observables}, we report analogous results for $4$-local 
operators 
and QFI computed with respect to a non-diagonal operator $\hat{T}$, defined 
in Eqs.~\eqref{e:4local} and~\eqref{e:spinflip}, respectively). 
We start by presenting the corresponding numerical ED
results which, as in the QFI case, display universal behavior in 
the disorder-averaged time series. Then, we use the Lindblad 
ME derived in Eq.~(\ref{e:me}) to further illuminate the universal dynamics and the salient features of the exact evolution. 
Finally, we present a spectral analysis of the corresponding Liouvillian, which 
explains the universality (initial state independence) within the 
Bourret--Markov approximation.

\subsection{Numerical results from exact diagonalization}
\label{s:subsec_syk4_partA}

With respect to the symmetry unbroken FH ground states, the expectation values 
of all the odd moments of the staggered magnetization operator $\hat{R}$ are 
zero.
Their ensemble averaged expectation values continue to show negligibly small 
fluctuations around zero during time evolution under the SYK$_{4}$ 
Hamiltonian.\footnote{One can prove formally that odd moments of $\hat{R}$ 
vanish as follows. The considered initial states, having zero magnetization, 
have 
spin-flip symmetry. In contrast, the staggered magnetization operator is odd 
under such a transformation. Regarding the SYK Hamiltonian, spin-flip (i.e., 
$2i\leftrightarrow 2i-1$) turns one disorder realization into a different one 
with the same probability of occurring. 
As a consequence, given an instance of the SYK Hamiltonian that produces a 
certain dynamics of $\hat{R}^k$, with $k$ odd, there will always exist another 
realization that generates dynamics of equal amplitude, but opposite sign, 
leading to a vanishing ensemble average.}
In contrast, the even moments exhibit the same super-exponential universal 
equilibration behavior as the QFI. 
This is illustrated in Fig.~\ref{f:fig_moments} for $k = 2$, $4$, and $6$ (higher-order moments are presented in Fig.~\ref{f:fig_moments_higher}). 
For visualization purposes, in the left column the expectation values of the 
operator moments are normalized to values $\leq\mathcal{O}(1)$ by an empirical 
factor 
\begin{equation}
	\mathcal{N}_{k}=N^{\left(\frac{3k}{4}-\frac{1}{2}\right)}.
 \label{e:scale_fac}
\end{equation}
The super-exponential approach to equilibrium is clearly visible in this data. 

As is evident from the right columns of Figs.~\ref{f:fig_moments} 
and~\ref{f:fig_moments_higher}, a rescaling using Eq.~\eqref{e:rescaling} 
collapses the even moments evolved from different initial states onto a single 
curve. 
During most of the evolution, this collapsed curve can be well approximated by 
a Gaussian. 
In the transient regime, curves corresponding to different 
initial states for even $k\geq 4$ do show small deviations, an effect that 
becomes 
more prominent for larger $N$. In other words, while the 
universality found for $k =2$ is very robust, for larger $k$ it becomes approximate in an 
intermediate time window.
Below, we describe this feature in detail via a spectral analysis of the 
Liouvillian. 
This finding also suggests that universality is more precise for few-body 
operators, as is further corroborated by comparison with the global many-body 
observable of the survival probability, see Fig.~\ref{f:fidelity_log_log}.
However, we do not exclude the possibility of a suitably 
constructed, highly non-local observable exhibiting universality with 
respect to Eq.~\eqref{e:rescaling}.

An interesting feature of the different moments is that their respective curves 
shift towards earlier times with increasing order $k$.
That is, the higher the moment order is, the faster is the equilibration.
Accordingly, Gaussian fits to the universal curves for $k=2,4,6$ yield the 
decreasing decay times $\tau = 1.52, 1.42, 1.35$, respectively, see 
Fig.~\ref{f:fig_moments}.
To illustrate this effect further, we show the rescaled curves for $k = 2, 4, ..., 12$ in 
Fig.~\ref{f:fig_moments2to12}, where it appears the curves converge with sufficiently 
high order.
Again, this trend can be explained based on a spectral analysis of the 
Liouvillian as a function of moment order 
(see Sec.~\ref{s:MEuniversality} and inset of Fig.~\ref{f:fig_moments2to12} in 
App.~\ref{app:higher_moments} for details).

\begin{figure}[ht!]
 \includegraphics[height = 5.7cm]{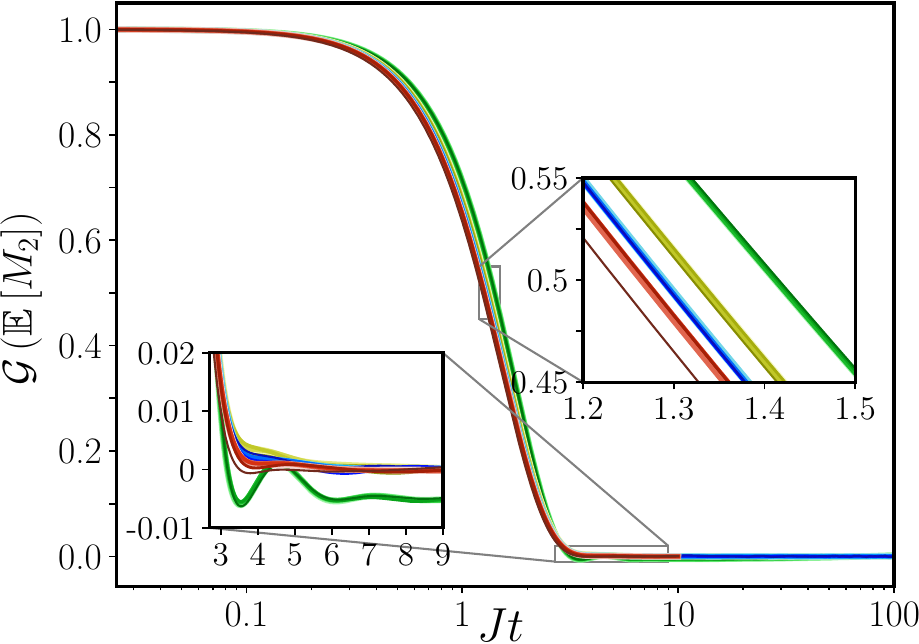}
 \caption{
      System-size dependency. Rescaled equilibration dynamics of the 
      disorder-averaged second moment, 
	  $\mathcal{G}\left(\mathbb{E}\left[M_{2}\right]\right)$, of the operator 
	  $\hat{R}$ under the complex SYK$_4$ Hamiltonian for $N = 8, 12, 16$, and 
	  $20$ (green, yellow, blue, and red curves, averaged over $90000$, $2700$, 
	  $400$, and $50$ disorder realizations, respectively). 
      Dark to light shadings of a given color correspond to the different 
      initial states of Fig.~\ref{f:fig_schem_univ_fq}.
    We remark that initial states correspond to the FH model with anti-periodic 
    boundary conditions for $N=8,16$, and with periodic ones for $N=12,20$. 
    Still, this does not affect the study of behavior with system size: initial 
    state independence is observed for all $N$, and the 
    universal curves equilibrate faster, with an indication of convergence with increasing $N$ to a 
    fastest decay curve (right inset). 
    The small spread of the curves for a given $N$ at 
    intermediate times (left inset) is of statistical nature due to finite 
    sample sizes.
}
 \label{f:fig_sys_size_dep}
\end{figure}

To study the finite-size dependency of the rescaled universal curves, we 
consider in Fig.~\ref{f:fig_sys_size_dep} the representative case $k = 2$ of 
the operator moments.
We consider the exact evolution for systems consisting of  $N = 8, 12, 16$, and $20$ 
complex fermionic modes, and investigate the dynamics in the number preserved sector of 
half filling, for which we employ a state-of-the-art, highly optimized exact 
diagonalization method. For the largest system size, the Hamiltonian matrix 
dimension is
$D=N!/[(N/2)!(N/2)!] = 184756$.
Due to the disorder, no symmetries other than particle number conservation can 
be used, and due to all-to-all connectivity the matrix is denser than for 
models with finite-range interactions. 
However, thanks to the self-averaging nature of the model~\cite{sonner_2017}, with 
increasing $N$ smaller numbers of disorder realizations suffice for satisfactory 
convergence~\cite{schiulaz_2020} (we consider $90000, 2700, 400$, and $50$ 
ensemble 
members for the above values of $N$).
We note that, as an alternative to exact diagonalization, semiclassical methods have also been shown to successfully capture the dynamics of operator expectation values in Ref.~\cite{Davidson_etal2017}.

With increasing $N$, faster equilibration as well as an approach to convergence is observed (see the right inset in Fig.~\ref{f:fig_sys_size_dep}, which highlights the dynamics in 
the transient time domain).
A similar feature has been seen in the initial dynamics of other quantities for time evolution under 
the SYK~\cite{cotler_2017} and other disordered, chaotic Hamiltonians~\cite{herrera_2019,
herrera_2020}, as well as random matrices~\cite{chenu_2019}.
The dependence on the system size can again be understood from the spectral analysis of the Liouvillian (see Sec.~\ref{s:MEuniversality}).
For smaller $N$, the curves show oscillations before equilibrating to the 
steady state value. In addition, for $N=8$, at large times the equilibrated 
curves slowly drift from the steady state value, with a rate that depends on 
the considered initial states. As a consequence, the rescaled curves cross 
zero and become negative at intermediate time, which is in accordance with 
Eq.~\eqref{e:rescaling} (see the left inset in Fig.~\ref{f:fig_sys_size_dep}, 
which highlights the approach of the curves to the steady state value). 
Both the transient oscillations and the drift can be attributed to finite-size 
effects, which become less prominent with increasing $N$. The oscillations may 
have random matrix 
origin, and are, in fact, also a feature of the SYK$_{2}$ model~\cite{bandyopadhyay_2021}. 
In contrast, the drift is due to finite-size induced non-exact uniformity in the 
distribution of initial states' amplitudes over the SYK$_{4}$ Hamiltonian 
spectrum. 
A similar drift has also been reported in the dynamics of the purity 
under Poissonian and Gaussian Unitary Ensemble random 
matrices~\cite{Kropf_etal16}.
The drift is reminiscent of the \emph{correlation hole}, 
which appears in the evolution of survival probability, inverse participation
ratio, and correlation functions~\cite{lezama_2021} of chaotic systems.
Alternatively, the slow drift constitutes the ``ramp" of the \emph{dip-ramp}
found in the dynamics of the spectral form-factor~\cite{cotler_2017} of SYK and
random matrix models. Both terms describe the same phenomenon, which 
arises due to the long-range rigidity in the Hamiltonian spectrum.
The depth of the correlation hole for equal time correlation functions is known
to be suppressed for larger system size~\cite{lezama_2021}. This, in another
way, substantiates that the drift seen for $N =8$ is a finite size effect for
the observables considered in the present study.

\begin{figure}[ht!]
 \includegraphics[height = 5.7cm]{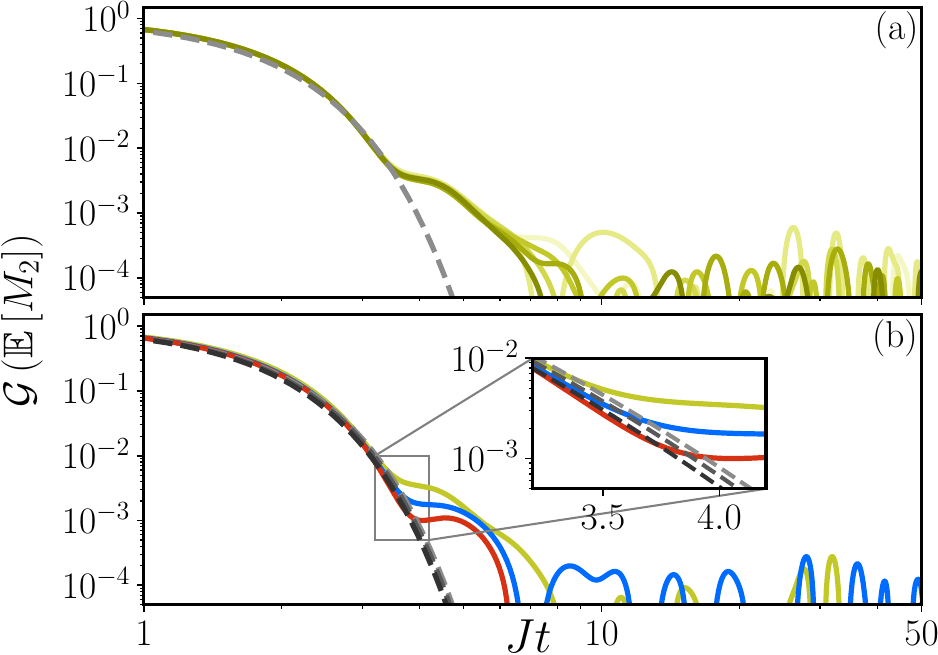}
 \caption{
	Universal dynamics, in logarithmic scales. 
	(a) Rescaled dynamics of $M_{2}$ of $\hat{R}$, defined in 
	Eq.~\eqref{e:stagmagSYK}, for $N = 12$ with 
	respect to different 
	initial states which are the FH ground states for $U/\mathcal{J} = 0, 2, 4, 
	6, 8,$ and $10$ 
	(light to dark shading). The curves for $U/\mathcal{J} = 2, 4, 6, 8$ 
	and $10$ are averaged over $10000$ disorder realizations, respectively. 
	For $U/\mathcal{J} = 0$ an ensemble of $100000$ 
	realizations is used due to slower convergence in disorder averaging.
	A significant part of the equilibration can be well approximated by a 
	Gaussian (gray dashed curve), followed by a marginal domain of an 
	approximate power-law behavior. 
	The universality holds throughout these regimes, and seems to extend to 
	a larger time domain with increasing disorder sample averaging. 
	(b) Rescaled curve for system sizes $N = 12$ (yellow), $16$ (blue), $20$ 
	(red), for the initial FH ground state at $U/\mathcal{J} = 4$. 
	With increasing $N$, the domain of power-law behavior seems to diminish 
	and the description by a Gaussian to improve (inset: the dashed, light 
	to dark gray, curves correspond to Gaussian fits for $N=12,16$ and $20$, 
	respectively). 
	}
 \label{f:fig_moment2_log_log}
\end{figure}

To scrutinize the universality in more detail, we depict in 
Fig.~\ref{f:fig_moment2_log_log}a the evolution of 
$\mathcal{G}\left(\mathbb{E}\left[M_{2}\right]\right)$ on logarithmic scales 
around the transient domain and at the verge of attaining the steady state 
(with $N=12$ and $10000$ disorder realizations). The curves show universality 
within the 
numerical precision, which until a time of about $Jt\approx 6$ corresponds to the 
thickness of the curves.
The vast part of the corresponding universal curve is well approximated by a Gaussian, 
which reaches well into intermediate times $1\lesssim Jt \lesssim 10$. 
This super-exponential behavior is followed by a marginal time domain 
with an indication of power-law decay to the steady-state. 
From Fig.~\ref{f:fig_moment2_log_log}a, one sees that the universality manifests 
throughout these regimes, while we cannot make predictions at larger times due to 
lacking convergence in disorder averaging.
In Fig.~\ref{f:fig_moment2_log_log}b, we show the rescaled curve for $N = 12, 
16$, 
and $20$ to illustrate the system size dependence (choosing the initial FH 
ground state for $U/\mathcal{J} = 4$ as a representative case). 
With increasing system size, the time-interval of the power-law decay seems to 
diminish, and the Gaussian appears to fit the curve for larger times (see 
inset). 
Closer to the steady state, a larger number of realizations is required for 
disorder averaging 
to converge.
We further observe that the noninteracting FH ground state ($U/\mathcal{J}=0$) 
requires a larger sample size for convergence than other initial states.

\subsection{Numerical results from master equation}\label{s:subsec_syk4}

In this section, we apply the open-system formalism of Sec.~\ref{s:master_eq} to 
the SYK$_4$ Hamiltonian. We explicitly show 
the form of the jump 
operators and dissipation rates that were used for the ME 
    simulations of the QFI presented in Fig.~\ref{f:qfi_meVSed} and for the 
    operator 
    moments of the preceding section. 
To do so, one simply needs to rewrite the Hamiltonian of Eq.~\eqref{e:Hsyk4} in 
the 
generic form of Eq.~\eqref{e:Hgeneric}, and then
read off the jump operators $\hat{h}_{l_\alpha}^{(\alpha)}$ and disorder 
functions $\xi^{(\alpha)}_{l_\alpha}$ that govern the dissipation rates as 
follows.

First, since Eq.~\eqref{e:Hsyk4} is a purely disordered 
Hamiltonian, we have $\hat{H}_0=0$ and the Liouvillian in Eq.~\eqref{e:me} 
generates purely dissipative dynamics.
Second, for the SYK$_{4}$ Hamiltonian, we have the multi-index $l_\alpha=i_1 
i_2;j_1 j_2$, and can identify three Hamiltonian terms $\hat{H}_\alpha$ with 
jump operators
\begin{equation}\label{e:jops_syk4}
\hat{h}_{l_\alpha}^{(\alpha)} = \begin{dcases}
\hat{c}_{i_1}^\dagger \hat{c}_{i_2}^\dagger \hat{c}_{i_1} \hat{c}_{i_2} &, \quad
\alpha 
= 1 \\
\hat{c}_{i_1}^\dagger \hat{c}_{i_2}^\dagger \hat{c}_{j_1} \hat{c}_{j_2} + 
\mathrm{H.c.} &, \quad\alpha = 2 \\
i \hat{c}_{i_1}^\dagger \hat{c}_{i_2}^\dagger \hat{c}_{j_1} \hat{c}_{j_2} + 
\mathrm{H.c.} &, \quad\alpha = 3
\end{dcases},
\end{equation}
and corresponding time-independent disorder coefficients
\begin{equation}\label{e:syk4_xi}
\xi_{l_\alpha}^{(\alpha)} = \begin{dcases}
4 J_{i_1 i_2;i_1 i_2} / (2N)^{3/2} &,\quad \alpha = 1\\
2 \mathrm{Re} J_{i_1 i_2;j_1 j_2} / (2N)^{3/2} &, \quad\alpha = 2 \\
2 \mathrm{Im} J_{i_1 i_2;j_1 j_2} / (2N)^{3/2} &,\quad \alpha = 3
\end{dcases}.
\end{equation}
Explicitly, for 
$\alpha=1$ the multi-indices are $l_1=i_1 i_2;i_1 i_2$ with $i_1 > i_2$, 
whereas for 
$\alpha=2,3$ the multi-indices are $l_\alpha = i_1 i_2;j_1 j_2$ with 
$i_1>i_2,j_1>j_2$ and $(i_1,i_2) \neq (j_1,j_2)$.
Finally, we use the above expressions to determine the dissipation 
rates of Eq.~\eqref{e:dissip}. The relevant time integral is trivial in this 
case, 
and the rates are given by
\begin{equation}\label{e:drates_syk4}
2 f^{(\alpha)}_{l_\alpha, k_\alpha}(t) = 2t \eave{ \xi_{l_\alpha}^{(\alpha)} 
\xi_{k_\alpha}^{(\alpha)} 
} =
2t \left(
16 J^2 / (2N)^{3} \delta_{l_\alpha, k_\alpha} \right),
\end{equation}
for $\alpha=1$. Similarly, for $\alpha=2,3$, the dissipation rates are 
proportional to $2t [4 (J^2/2) / (2N)^{3} ]$, and some care must be taken as 
there exist 
pairs of indices $l_\alpha 
\neq k_\alpha$ for which $\eave{ \xi_{l_\alpha}^{(\alpha)} 
\xi_{k_\alpha}^{(\alpha)} } \neq 0$ (the origin of these correlations is 
explained in App.~\ref{app:MEapp}). 

The main point is that the time-independent disorder correlations of 
the SYK$_4$ model---or indeed any SYK$_q$ model given by Eq.~\eqref{e:Hsykq}---yield 
dissipation rates in the Bourret--Markov ME that grow linearly in time. This 
property yields the super-exponential approach to 
equilibrium, 
as already discussed in Sec.~\ref{s:master_eq} in the context of the QFI. 

\begin{figure}[ht!]
    \includegraphics[width = 
    \linewidth]{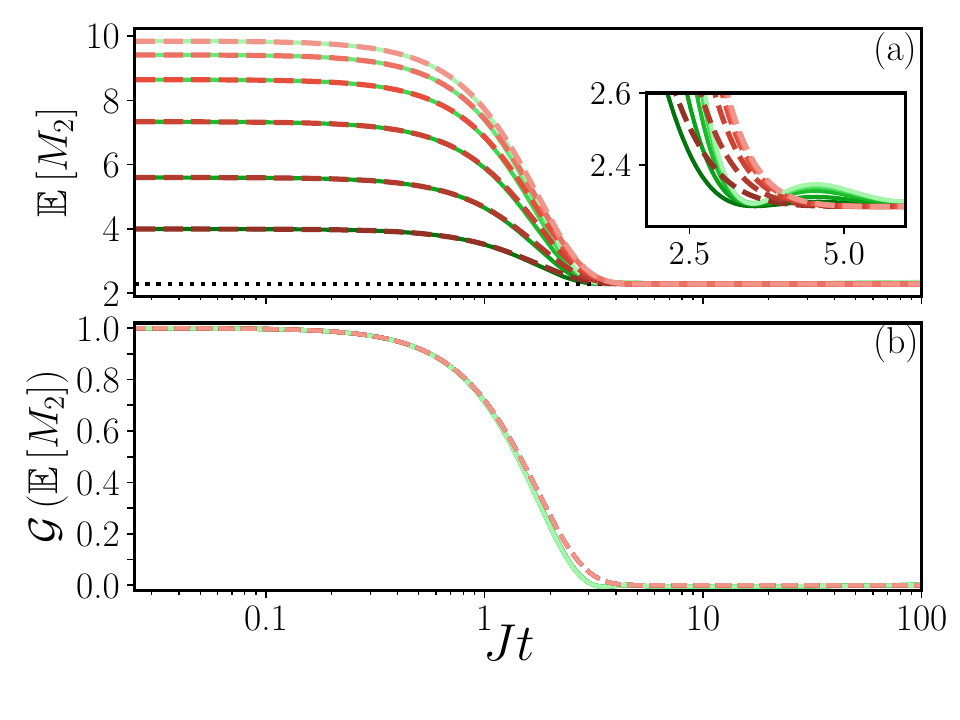}
    \caption{
        Comparison of ED and ME results for $M_2$ of $\hat{R}$, with $N=8$. 
        Colors and shading of ED (solid) and ME (dashed) curves are as in 
        Fig.~\ref{f:qfi_meVSed}.
        The analytically predicted steady-state value of 
        Eq.~\eqref{e:ssvalueME} 
        for the half-filling sector is given by the black dotted line.  
        (a) For each initial state, the ME simulation reproduces the exact 
        dynamics very well. 
        There is a discrepancy at intermediate times due to 
        non-Markovian effects not captured by the ME (inset). 
        (b) The ME reproduces the collapse to a universal curve under rescaling 
        \eqref{e:longtimeav}.
    }
    \label{f:ed_vs_me_syk4}
\end{figure}

Figure~\ref{f:ed_vs_me_syk4} shows a comparison of ME and ED simulations 
for moment $M_2$ of operator $\hat{R}$ defined in Eq.~\eqref{e:stagmagSYK}. 
As for the QFI, the super-exponential approach to equilibrium 
is captured, as well as the early and late time dynamics. 
We again observe a discrepancy between ED and ME simulations at intermediate times, a 
signature of correlation and memory effects as discussed in 
Sec.~\ref{s:master_eq}.
As mentioned above, the ME curves contain no fit parameters. 
Recalling that hermitian jump operators guarantee the infinite temperature 
ensemble to be a (not necessarily unique) steady state of the general 
Liouvillian in Eq.~\eqref{e:me}, we determine the infinite temperature 
steady-state value of 
$M_2$ within the half-filling sector $N=2Q$ to be 
\begin{equation}\label{e:ssvalueME}
\tr \left( \hat{R}^2(t) \hat{\rho}_\infty \right)  = \left( \frac{N}{2 
    \sqrt{N-1} } 
\right)^2 .
\end{equation}
As Fig.~\ref{f:ed_vs_me_syk4} shows, this value 
agrees well with the steady-state plateaus of the exact unitary dynamics averaged 
over disorder realizations.

\subsection{Analyses of Universality in ME formalism}
\label{s:MEuniversality}

Having used the ME framework to formally demonstrate the 
Gaussian decay via Eq.~\eqref{e:formal_time_evo}, we now study the origins of 
the observed universality within this formalism.
To this end, we numerically obtain the spectrum and eigenmodes of the SYK$_4$   
Liouvillian superoperator via a matrix representation thereof (see, e.g., 
Refs.~\cite{Minganti18,Lidar19,RivasHuelga} for detailed descriptions of such 
a procedure and for spectral properties of Liouvillian superoperators). 
As we will see, the population of 
various initial states and observables in the corresponding eigenspaces
conspires to produce a universal curve under the rescaling 
$\mathcal{G}$ defined in Eq.~\eqref{e:rescaling}.

In general, a superoperator 
$\mathcal{L}$ is not normal, and thus has distinct left and right eigenmodes. 
However, as a result of $\hat{H}_0=0$, $(\hat{h}^{(\alpha)}_{l_\alpha})^\dagger 
= 
\hat{h}^{(\alpha)}_{l_\alpha}$, and 
$F^{(\alpha)}_{l_\alpha,k_\alpha} = F^{(\alpha)}_{k_\alpha,l_\alpha} \in \RR$, 
the 
Liouvillian of the SYK model $\mathcal{L}(t)=2t\mathcal{D}$ is Hermitian and 
thus normal, as made explicit by the matrix representation given 
in Eq.~\eqref{e:Lvec}. Therefore, the 
left and right eigenmodes of $\mathcal{D}$ coincide, and one can always 
form a Hermitian basis for each eigenspace. We use the index $i\geq0$ to label these 
eigenspaces, which in general have a $d_i$-fold degeneracy. The $d_i$ Hermitian 
eigenmodes within the $i$th eigenspace are denoted as 
$\hat{\rho}_{i,\alpha_i}$, where $\alpha_i=1,\ldots,d_i$. The eigenmodes are 
orthogonal with respect to the Hilbert--Schmidt norm 
$\tr(\hat{\rho}_{i,\alpha_i}^\dagger \hat{\rho}_{j,\alpha_j}) = \delta_{i,j} 
\delta_{\alpha_i,\alpha_j}$ 
and thus form a basis of $\mathcal{B}(\mathcal{H})$, the space of linear 
operators acting on $\mathcal{H}$. 
For all (in our case typically degenerate) eigenspaces $i$, the corresponding 
eigenvalue $\lambda_i$ is real and 
negative. So, we order the eigenspaces according to the magnitude of their 
respective eigenvalues as $|\lambda_0| < |\lambda_1| < \ldots$ . 

We can decompose any initial state and observable in 
$\mathcal{B}(\mathcal{H})$, 
respectively, as
\begin{equation}\label{e:decompositions}
\hat{\rho}(0) = \sum_{i\geq 0} \sum_{\alpha_i=1}^{d_i} c_{i, \alpha_i} 
\hat{\rho}_{i, 
\alpha_i}  \,\,\text{  and  }\,\,
\hat{O} = \sum_{i\geq 0} \sum_{\alpha_i=1}^{d_i} o_{i, \alpha_i} \hat{\rho}_{i, 
\alpha_i} ,
\end{equation}
with real coefficients $c_{i, \alpha_i}=\tr ( \hat{\rho}(0) \hat{\rho}_{i, 
\alpha_i} )$ and $o_{i, \alpha_i}=\tr ( \hat{O} \hat{\rho}_{i, \alpha_i} )$.
It then follows that any state time-evolved under the SYK$_4$ dissipator 
according to Eq.~\eqref{e:formal_time_evo} is given by 
\begin{equation}
\tilde{\rho}(t) = e^{t^2\mathcal{D}} \hat{\rho}(0) = \sum_{i\geq 0} e^{-t^2 
    \abs{\lambda_i} } \sum_{\alpha_i=1}^{d_i} c_{i, \alpha_i} \hat{\rho}_{i, 
    \alpha_i} .
\end{equation}
Since Liouvillian dynamics are trace preserving, we have $\lambda_0 = 0$. 
Consequently, $\lim_{t\to\infty} \tilde{\rho}(t) $ is given 
in terms of the eigenmodes corresponding to $\lambda_0$ \cite{Minganti18}. 
For the Liouvillian of the SYK$_4$ model, we find $\lambda_0$ to be 
non-degenerate, implying a unique steady-state in the present case of study, in 
agreement with rather general conditions \cite{Nigro2019}.
The Liouvillian spectrum $\{\lambda_i\}$ sets the time-scales of the dynamics 
of any observable quantity. 

To analyze the universality of operator moments and the pure-state QFI observed 
under the rescaling $\mathcal{G}$ defined in Eq.~\eqref{e:rescaling}, we 
numerically obtain $\{\lambda_i\}$ and $\{\hat{\rho}_{i, \alpha_i}\}$ from an 
exact diagonalization of a matrix representation of $\mathcal{D}$ (see 
App.~\ref{app:MEapp} for details).
For sufficiently large $Jt_0$ [see Eq.~\eqref{e:longtimeav}], the 
long-time average in $\mathcal{G}$ is simply the contribution due to the 
steady-state 
eigenmode $\hat{\rho}_0$. 
With the above eigendecompositions in hand, we can thus express 
any rescaled operator expectation value as
\begin{equation}\label{e:universal_curve}
\mathcal{G}\bigl( \mathrm{tr}\bigl( \hat{O} \tilde{\rho}(t) \bigr) \bigr) = 
\frac{ 
    \sum_{i\geq 1} e^{-t^2 \abs{\lambda_i} } A_i }{ \sum_{i \geq 1} A_i },
\end{equation}
where $A_i = \sum_{\alpha_i=1}^{d_i} c_{i, \alpha_i}  o_{i, \alpha_i}$ is the 
effective amplitude within the $i$th eigenspace.
Universality across different initial states can occur if:
\begin{enumerate}[(i)]
    \item \label{l:universal_single} The observable and initial state 
    decompositions of Eq.~\eqref{e:decompositions} intersect in only one and 
    the same eigenspace $i^*>0$ for all initial states. 
    \item \label{l:universal_multiple} The decompositions intersect in multiple 
    degenerate eigenspaces, but the $ c_{i, \alpha_i}  o_{i, \alpha_i}$ are 
    distributed symmetrically about $0$ in all but one eigenspace $i^*>0$.
\end{enumerate}
In both cases, there exists only one non-zero amplitude $A_{i^*}$, and $A_i = 0$ $\forall i\neq i^*$, such that Eq.~\eqref{e:universal_curve} reduces 
	to the same Gaussian curve $\mathcal{G}\bigl(\bigl\langle \hat{O}(t) 
	\bigr\rangle \bigr)=e^{-t^2 \abs{\lambda_{i^*}}}$ for all initial states.

\begin{figure}[ht!]
    \includegraphics[width =   
    \linewidth]{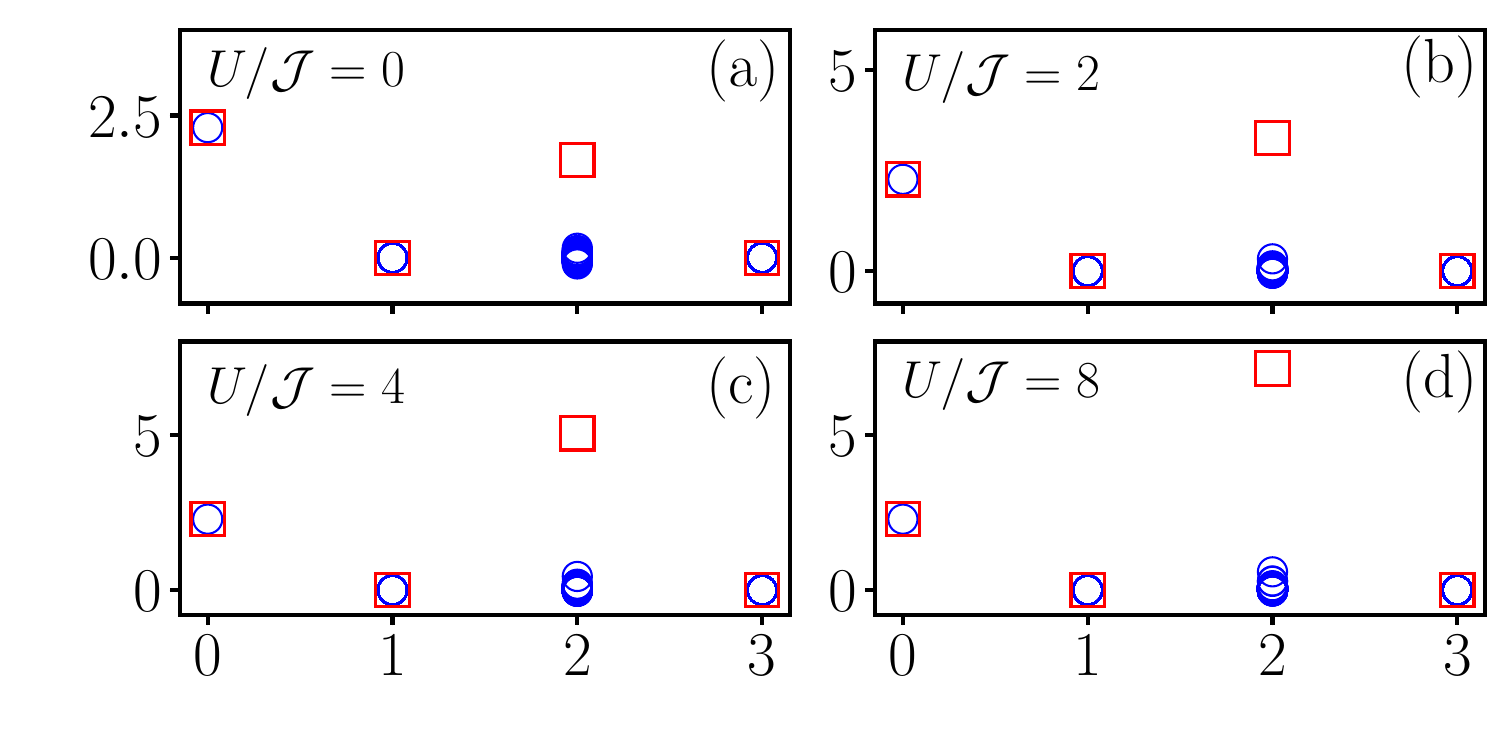}
    \caption{
        Distributions of $c_{i,\alpha_i} o_{i,\alpha_i}$ (blue circles) and 
        $A_i$ (red squares) of Eq.~\eqref{e:universal_curve} for 
        $M_2$ of $\hat{R}$, for different initial FH ground states with 
        $U/\mathcal{J}=0,2,4$ and $8$. Horizontal axes indicate 
        the eigenspace index of Table~\ref{t:syk4_spectrum}. 
        For all initial states, only one eigenspace $i^*=2$ has non-zero 
        effective amplitude $A_2$, yielding the universal 
        evolution \mbox{$\mathcal{G}[M_2(t)] = e^{-t^2 \abs{\lambda_2}}$}.
    }
    \label{f:liouvillian_eigenspaces_states}
\end{figure}

We apply the above criteria to the first four moments 
$M_k$ of the staggered magnetization $\hat{R}$ [see Eq.~\eqref{e:stagmagSYK}] 
in the SYK$_4$ model.
Table~\ref{t:syk4_spectrum} lists the spectrum of $\mathcal{D}$ for a system of 
$N=8$ modes at half filling.
The corresponding distributions of $c_{i,\alpha_i} o_{i,\alpha_i}$ and $A_i$  
for $M_2$ are shown in Fig.~\ref{f:liouvillian_eigenspaces_states} for  
different initial FH ground states.
Since---for $i>0$---$c_{i,\alpha_i} o_{i,\alpha_i} \neq 0$ 
only for $i=2$, universality of 
type (\ref{l:universal_single}) for $M_2$ follows immediately with $i^*=2$. 
The reason why this occurs lies in the decomposition of the observable: 
We observe $o_{i,\alpha_i}=0$ for $i>0$ and $i\neq 2$. 
This demonstrates why the choice of the initial state is completely 
irrelevant---regardless of the values of $c_{i,\alpha_i}$, there can be only 
a single $A_i\neq 0$ for $i>0$.

\begin{figure}[ht!]
    \includegraphics[width = 0.95\linewidth]
     {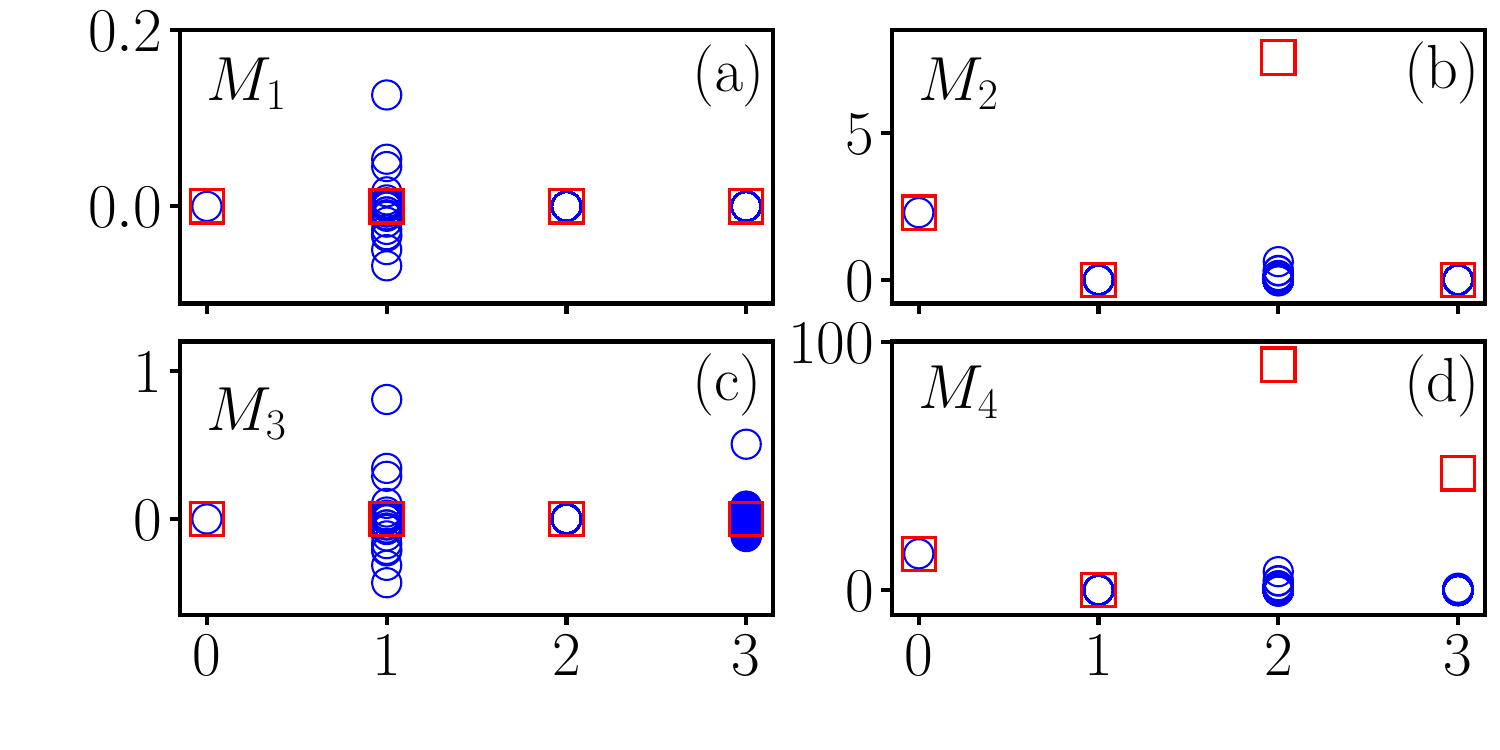}
    \caption{
        Similar to Fig.~\ref{f:liouvillian_eigenspaces_states}, 
        but for different moments $M_1,M_2,M_3,M_4$ of $\hat{R}$, for the  
        initial FH ground state at $U/\mathcal{J}=10$. 
        For the highest moment, occupation of two eigenspaces can be observed, 
        indicating that universality deteriorates in many-body operators.
    }
    \label{f:liouvillian_eigenspaces_moments}
\end{figure}

In Fig.~\ref{f:liouvillian_eigenspaces_moments}, we display the behavior of 
moments $M_1,M_2,M_3,M_4$ for the FH ground state at $U/\mathcal{J}=10$. 
For odd moments, we find $A_i=0 \,\, \forall i$, making them trivially universal: 
For $M_1$, only $c_{1,\alpha_1} o_{1,\alpha_1} \neq 0$, whilst for $M_3$ 
additionally $c_{3,\alpha_3} o_{3,\alpha_3} \neq 0$. In either case, these 
terms are distributed near symmetrically about $0$, such that the effective 
amplitudes vanish (for more details see Sec.~\ref{s:subsec_syk4_partA}).

In contrast, even moments have non-zero effective amplitudes in at least one 
eigenspace 
besides that of the steady state.
In analogy to Fig.~\ref{f:liouvillian_eigenspaces_moments}, we have verified for
a range of initial FH ground states, as well as others such as the 
Neel state, that for a given even moment, any non-zero amplitudes $A_i$ always 
occupy the same 
eigenspaces. Concretely, for $k\geq 4$, we find the same two non-zero 
effective amplitudes $A_2, A_3$, yielding an approximately universal 
super-exponential decay $\mathcal{G}[M_{k}(t)] = (A_2 e^{-t^2 \abs{\lambda_2}} 
+ A_3 e^{-t^2 \abs{\lambda_3}}) / (A_2 + A_3)$ for even integers $k\geq4$. 

In summary, we find (i) odd moments vanish for all 
$t$ (and are thus trivially universal), (ii) the second moment exhibits truly 
universal 
super-exponential decay as is shown in 
Fig.~\ref{f:liouvillian_eigenspaces_states}, and (iii) higher-order even 
moments display approximate universality. 
As this analysis shows, the Bourret--Markov ME reproduces the universal 
features observed in our exact numerics.

To conclude this section, we study the dependence of the Liouvillian 
spectrum on the system size $N$. 
We find that all non-zero eigenvalues decrease as $N$ is increased from $6$ to 
$8$.
In particular, for $\lambda_2$---the only timescale entering $M_2$ (see 
Fig.~\ref{f:liouvillian_eigenspaces_states})---we find, respectively, the 
values $-0.28$ and $-0.33$.
This explains the shift to a faster equilibration time of 
$M_2$ with increasing $N$, as observed previously in  
Fig.~\ref{f:fig_sys_size_dep}.
Note that this shift of $M_2$ exhibits a convergence, i.e., decreases as $N$ is 
increased.
We thus expect the eigenvalues $\lambda_i$ to not decrease indefinitely with 
$N$, but to individually approach some asymptotic value (for SYK$_{q=2}$ this 
behavior can be shown analytically, but remains to be shown for $q\geq 4$ 
\cite{bandyopadhyay_2021}).
However, the enlarged Hilbert-space, inherent to the process of mapping 
$\mathcal{D}$ to the matrix form of 
Eq.~\eqref{e:Lvec}, limits our study of 
the Liouvillian spectrum to $N\leq 8$, thus preventing us from probing 
this convergence.

In this context, we emphasize that the aim of the proposed ME framework is not 
to provide an efficient way of simulating the disorder-averaged dynamics.
Rather, the aim is to gain additional insights by establishing a theoretical 
mapping to an open quantum system.
For example, our present study reveals a non-trivial highly-degenerate 
eigenspace structure of the effective Liouvillian superoperator, through which 
we can explain the observed universality.
The origin of this eigenspace structure is, briefly put, an operator size symmetry where each degenerate eigensector
    corresponds to a set of operators sharing a common operator size. For an in-depth study of this structure, we refer
     the interested reader to our follow-up work Ref.~\cite{Paviglianiti_etal2022}.
%
\begin{table}[ht]
    \caption{ \label{t:syk4_spectrum}
        Spectrum of SYK$_4$ Liouvillian, obtained
        by exact diagonalization of the matrix representation given in
        Eq.~\eqref{e:Lvec}, for $N=8$ fermionic modes at half filling.
    }
    \begin{tabularx}{\linewidth}{ccc}
            \hline\hline
            Eigenspace index $i$ & Eigenvalue $\lambda_i$ & Degeneracy $d_i$\\
            \hline
            0 &  0.000000 & 1 \\
            1 &  -0.234375 & 63 \\
            2 &  -0.328125 & 720 \\
            3 &  -0.351562 & 4116 \\
            \hline\hline
    \end{tabularx}
\end{table}

\section{Conclusion and Discussion}
\label{s:conclusion}

In summary, we have theoretically investigated post quench equilibration 
dynamics of a system of randomly interacting fermions described by the complex 
SYK$_4$ model. 
By numerically studying the disorder-averaged exact evolution of a set of local 
observables and their higher-order moments, we find that the equilibration 
process is universal.
The curves illustrating the equilibration of different initial states overlap 
throughout the dynamics under a straightforward rescaling, revealing
the independence of the dynamics on chosen initial states.
The equilibrated steady state, which is the Gibbs infinite temperature state in 
the present study, is reached in fast time-scales of leading-order processes 
determined by the variance of the disordered interaction.
In addition, the universal equilibration curve can be well approximated by a 
Gaussian, yielding fast super-exponential equilibration dynamics.

In order to achieve an analytical understanding of the numerical findings, we have formulated a 
theoretical framework based on the Novikov--Furutsu theorem. 
This framework describes how a disordered quantum many-body system undergoes an 
effective dissipative dynamics due to phase mixing in the ensemble averaged 
evolution rather than to interactions with a heat bath 
\cite{vanKampen1974,Kropf_etal16}. 
Thanks to the generality of the formulation, its scope for applications extends 
beyond the present investigation. 
Employing Bourret and Markov approximations, we obtain a Lindblad master 
equation that successfully describes the key features of the observed super-exponential 
equilibration dynamics under the SYK$_4$ model.
Furthermore, a spectral analysis of the corresponding Liouvillian superoperator 
illuminates 
the exact universality of low-order moments, representing few-body observables, 
as well as an approximate universality of higher-order moments representing 
many-body observables.

The Novikov--Furutsu theorem has been used extensively in the literature for the study of 
systems with noise of short correlation 
time~\cite{Montiel_etal19, Ancheyta_etal2021, Benatti_etal2012}, and equations equivalent to those derived in Sec.~\ref{s:master_eq} have been obtained to second order in perturbative noise strength  
\cite{Chenu_etal2017, budini_2000, budini_2001, Bourret1962, Dubkov1977}.
In the present scenario, where noise correlation times are formally infinite 
and where the disordered interactions provide the dominant (because only) 
energy scale, there is at first sight no reason for such perturbative 
approaches to hold. 
Yet, the strong chaoticity of the SYK$_{4}$ model leads to a fast decorrelation, making 
the Bourret--Markov approximation an excellent description of the exact dynamics.

The salient features of the universal curve occur on sizable absolute 
scales and very fast time scales, on the order of the mean interaction 
strength, and they can be extracted from the observation of local observables 
following a simple global quench. Thus, the discussed effects should be readily 
observable in forthcoming laboratory implementations of the SYK model, for 
which several proposals have recently been put forward 
\cite{danshita_2017, pikulin_2017, chew_2017, chen_2018, GarciaAlvarez_etal2017, 
bentsen_2019_2, WeiSedrakyan_2021}.
Moreover, the mapping to the purely dissipative Lindblad equation may also open 
new ways for simulating SYK matter using engineered open quantum dynamics.

We hope our findings will also stimulate further theoretical investigations to 
obtain and understand universality in out-of-equilibrium dynamics of other 
quantum many-body systems. 
A topic of our immediate future investigation is, e.g., in how far the universality survives in disordered models without all-to-all connectivity. 
Further, while we have considered in this work only the SYK$_4$ model, our 
findings hold equally well for other SYK$_q$ models~\cite{bandyopadhyay_2021}.
In the future, it will also be interesting to apply our master equation 
approach to other disordered models. 
Whilst time-independent Gaussian disorder will always yield a Gaussian decay according to our formalism, the universality (initial state independence) would depend on details of the Hamiltonian.
In particular, in models with a clean 
contribution, $\hat{H}_0\neq 0$, where one can expect a complex interplay 
between disorder and clean dynamics, such as the many-body localization 
transition~\cite{nandkishore_2015, Sierant_etal2017, abanin_2019, 
SierantZakrzewski_2021}. 
Various mathematical extensions of the presented master equation framework are 
also possible, e.g., to treat non-Gaussian disorder \cite{KlyatskinTatarskii72, 
haenggi_1978}. 
Finally, as this framework can naturally include dephasing noise with arbitrary 
correlation spectrum, it permits one to estimate the interplay between 
dissipation due to an external environment and dephasing due to disorder 
averaging, which is central, e.g., for environmental assisted quantum transport 
\cite{Plenio_2008, Rebentrost_2009, deLeonMontiel_etal2015, Maier_etal2019}.

\begin{acknowledgments}
We gratefully acknowledge useful discussions with Jean-Philippe Brantut, Julian 
Sonner, and Ricardo Costa de Almeida.
We acknowledge CINECA HPC project ISCRA Class C: ISSYK-HP10CE3PVN and 
ISSYK-2 (HP10CP8XXF).
We acknowledge support by Provincia Autonoma di Trento and the Google Research 
Scholar Award ProGauge.
This project has received funding from the European Research Council (ERC) 
under the European Union’s Horizon 2020 research and innovation programme 
(grant agreement No 804305).
This project has benefited from Q@TN, the joint lab between University of 
Trento, FBK-Fondazione Bruno Kessler, INFN- National Institute for Nuclear 
Physics and CNR- National Research Council.

\end{acknowledgments}

\appendix

\section{Details of SYK$_{q}$ model}
\label{app:DetailsOnSYKModel}
  
Here, we present the Hamiltonian of the general SYK$_{q}$ model of which 
Eq.~(\ref{e:Hsyk4}) is a special case with $q = 4$.
The SYK$_{q}$ Hamiltonian is governed by disordered all-to-all $(q/2)$-body 
interactions, and reads~\cite{gu_2020,baldwin_2020}
\begin{equation}
\begin{split}
&\hat{H}_{\mathrm{SYK}_{q}} =\\
&\frac{\mathcal{K}}{(q/2)!^{2}}
\sum_{\substack{i_1,...,i_{q/2} = 1 \\ j_1,...,j_{q/2} = 1}}^{N}
J_{i_{1}...i_{q/2};j_1...j_{q/2}} 
\hat{c}^{\dagger}_{i_{1}}...\hat{c}^{\dagger}_{i_{q/2}} 
\hat{c}_{j_{1}}...\hat{c}_{j_{q/2}},
\label{e:Hsykq}
\end{split}
\end{equation}
where $\mathcal{K}= \sqrt{[(q/2)!(q/2-1)!]/N^{q-1}}$ ensures the extensivity 
of the 
model. 
The interaction strengths in Eq.~\eqref{e:Hsykq} are complex Gaussian random 
variables, 
i.e., the real and imaginary parts of 
$\{J_{i_{1}...i_{q/2};j_1...j_{q/2}}\}$ are independent and normally 
distributed with variances parameterized by $J\in \RR_{>0}$ as
\begin{align}\label{e:syk_statistics}
  \eave{\Big(\Re(J_{i_{1}...i_{q/2};j_1...j_{q/2}})\Big)^2} = & \begin{cases} J^2,   
	  \quad \text{if } i_{l} =j_{l}, \forall \, l  \\ J^2/2, 
          \quad \text{otherwise,} \end{cases} \nonumber \\
  \eave{\Big(\Im(J_{i_{1}...i_{q/2};j_1...j_{q/2}})\Big)^2} = & \begin{cases}
0,     \quad \text{if } i_{l} =j_{l}, \forall \, l  \\
 J^2/2  \quad \text{otherwise.} 
\end{cases} 
\end{align}
Furthermore, the interaction strengths satisfy
\begin{equation}\label{e:asymmetry}
    \begin{split}
    J_{i_{1}...i_{q/2};j_1...j_{q/2}} &= J_{j_1...j_{q/2};i_{1}...i_{q/2}}^*,
    \\
    J_{i_{1}...i_{q/2};j_1...j_{q/2}} &= \mathrm{sgn}(\mathcal{P})
    \mathrm{sgn}(\mathcal{P}^{\prime}) J_{\mathcal{P}\{i_{1}...i_{q/2}\};
        \mathcal{P}^{\prime}\{j_1...j_{q/2}\}},
    \end{split}
\end{equation}
where $\mathcal{P}$ and $\mathcal{P}^{\prime}$ perform permutations of the 
indices, and 
$ \mathrm{sgn}(\mathcal{P}),\mathrm{sgn}(\mathcal{P}^{\prime})= \pm 1$ denote 
the sign of the permutations.
The first equality ensures 
Hermiticity of the SYK$_{q}$ Hamiltonian, whereas the second is due to the fermionic 
anticommutation relations of the creation and annihilation operators in 
Eq.~(\ref{e:Hsykq}).

\section{Higher-order moments}
\label{app:higher_moments}
 
\begin{figure}[ht!]
 \includegraphics[height = 6.0cm]{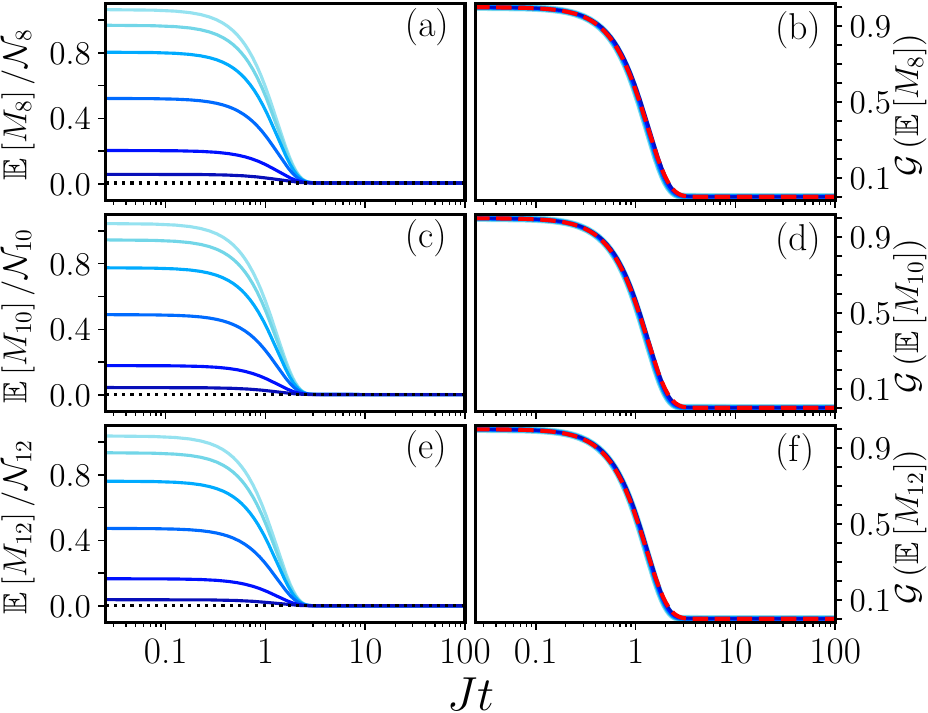} 
	\caption{
       Disorder-averaged equilibration dynamics of moments $k=8,10,$ and $12$ 
       of the operator $\hat{R}$ under the SYK$_4$ Hamiltonian for $Q=8$ 
       fermions occupying $N = 16$ modes. 
       Left column: Dynamics of $M_{8}$, $M_{10}$, and $M_{12}$ averaged over 
	   $400$ disorder samples. 
	   Right column: Corresponding dynamics rescaled by the function 
	   $\mathcal{G}$ given in Eq.~(\ref{e:rescaling}). 
	   As in the main text, the  dark to light shading of the curves 
	   represents initial FH ground states for $U/\mathcal{J} = 0, 2, 4, 6, 8$, 
	   and $10$, respectively.
   	   The dotted black lines mark the values of the operator moments calculated 
	   with respect to the Gibbs infinite temperature state.
       The rescaled curves are well approximated by Gaussian fits, 
       $\mathrm{exp}\left[-(Jt/\tau)^2\right]$, with $\tau = 1.32, 1.3, 1.28$ 
       for increasing $k$, respectively (dashed red curves).
	}
 \label{f:fig_moments_higher}
\end{figure}

\begin{figure}[ht!]
 \includegraphics[height = 6.0cm]{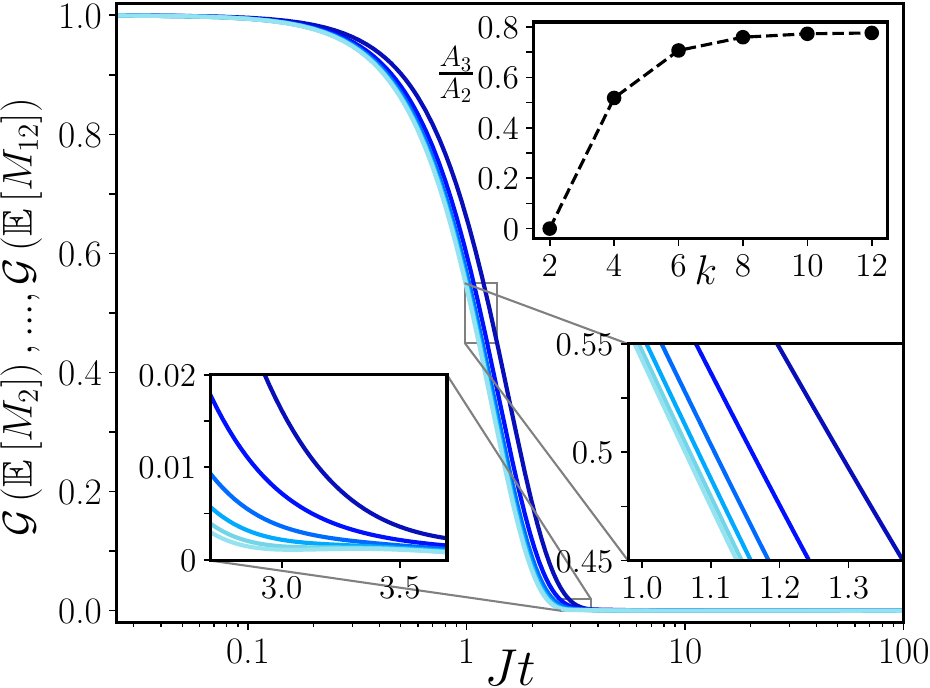} 
 \caption{
	  Rescaled dynamics of the disorder-averaged moments,
      $\mathcal{G}\left(\mathbb{E}\left[M_{k}\right]\right)$, of the operator
      $\hat{R}$ under the complex SYK$_4$ Hamiltonian for $N=16$ modes. 
      For all curves the 
      initial state is $U/\mathcal{J}=10$, and the dark to light 
      shading corresponds to different moment orders $k = 2, 4, 6, 8, 10$ and 
      $12$, 
      respectively.
      A shift to earlier times with increasing $k$ is evident, and is 
      further highlighted by the two lower insets.
      There is also an indication of convergence to a fastest decay time with 
      increasing $k$.
      This is further studied in the top inset, which shows the ratio $A_3/A_2$ 
      of the effective amplitudes of the $k$th moments in accordance with 
      Eq.~\eqref{e:universal_curve}. The monotonic increase, and indication of 
      saturation, with increasing $k$ shows that the faster time-scale $\abs{ 
      \lambda_3}$ is favored with increasing moment $k$.
      }
 \label{f:fig_moments2to12}
\end{figure}

Here, we present additional results for the quench dynamics of moments 
$k=8,10,12$ (see Fig.~\ref{f:fig_moments_higher}) of the staggered 
magnetization $\hat{R}$ defined in Eq.~\eqref{e:stagmagSYK}.
The rescaled dynamics (right column) show approximately universal Gaussian equilibration 
similar to the other moments $k<8$ studied in the main text.
However, unlike the case of $k =2$, the universality is not exact for these 
higher-order moments, as is explained in detail in Sec.~\ref{s:MEuniversality}.
In addition, the rescaled curves shift towards earlier 
time with 
increasing $k$, which is evident from the lower insets in 
Fig.~\ref{f:fig_moments2to12}. 
Further, the rescaled curves seem to converge at sufficiently large 
$k$. In order to 
explain this characteristic, we plot the ratio $A_{3}/A_{2}$ of the effective 
amplitudes [defined below 
Eq.~(\ref{e:universal_curve})] between the occupied eigensectors of the Liouvillian (see top inset of Fig.~\ref{f:fig_moments2to12}).
The ratio shows a monotonic increase and a saturation with respect to $k$. 
Together with the fact that $|\lambda_{3}|>|\lambda_{2}|$ (see 
Table~\ref{t:syk4_spectrum}), this explains---in accordance with  
Eq.~(\ref{e:universal_curve})---the shift of the rescaled curve to 
earlier times, as well as its convergence, with moment order.
We note that this convergence with moment order, and its explanation in 
terms of effective amplitude ratios, holds also for other 
choices of observables and/or initial states (not reported here), though the 
direction of convergence need not 
always be to earlier times but can also be to later times.

\begin{figure}[ht!]
    \includegraphics[height = 3.5cm]{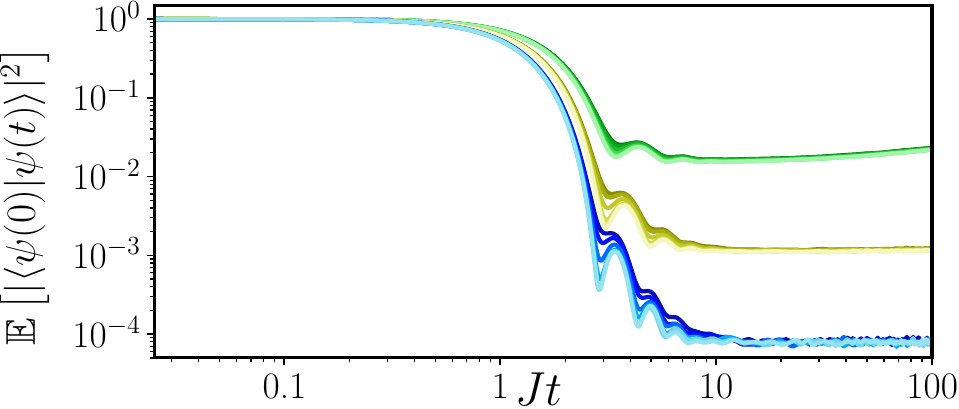}
    \caption{
        Ensemble averaged survival probability for system sizes $N=8$ 
        (green), $12$ (yellow), $16$ (blue), and sample sizes $90000$, $10000$, 
        $400$, respectively. Shadings of a given color represent 
        different initial states, as in 
        Fig.~\ref{f:fig_moments_higher}. 
	There is a clear lack of universality with respect to different initial 
	states from early to late times. The early-time Gaussian decay followed 
	by oscillations at intermediate times are 
        expected from random matrix theory (RMT)~\cite{herrera_2019,tavora_2016}.
    In contrast to RMT where universality is expected for the survival probability, here we observe an initial state \emph{dependence}, which can be attributed to the structure of the SYK Hamiltonian, imposed by the fermionic statistics.}
    \label{f:fidelity_log_log}
\end{figure}

Finally, we comment on the lack of exact universality observed for 
moments with $k>2$: The $k$th moment of $\hat{R}$ contains interactions of up 
to 
$k$ fermionic modes, and thus probes increasingly non-local physics for larger 
values of $k$.
As a limiting case of a truly global many-body quantity, we show in 
Fig.~\ref{f:fidelity_log_log} the disorder-averaged survival probability 
(or fidelity)
$\eave{ \abs{\langle \psi(0) \vert \psi(t) \rangle}^2 }$. 
For this quantity, the curves corresponding to different initial states do 
not collapse. Whilst this suggests that universality is absent for 
(superpositions of) highly non-local observables, we do not exclude the 
possibility that one may be able to specifically construct such an observable 
which does exhibit universality.
Finally, we note that the dynamics of the survival probability, namely an initial Gaussian 
decay followed by oscillations at intermediate times, is in accordance with the 
well established behavior of the post-quench dynamics of the survival 
probability within random matrix theory \cite{herrera_2019,tavora_2016}. 

\section{Additional observables}
\label{app:other_observables}
 Here, we extend our investigations to other observables in order to show the generality 
of our key findings. 

\begin{figure}[ht!]
 \includegraphics[height = 5.7cm]{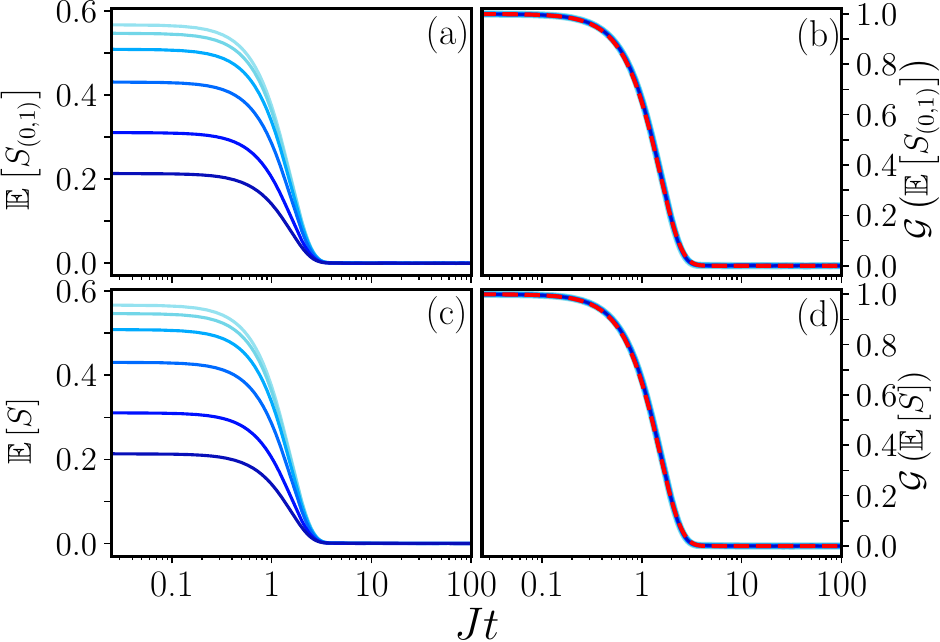} 
 \caption{Universal equilibration dynamics of the $4$-local operators 
          defined in Eq.~({\ref{e:4local}}), averaged over $400$ disorder 
          realizations of the complex SYK$_{4}$ Hamiltonian for $Q=8$ fermions 
          occupying $N = 16$ fermionic modes. Dark to light shading  
          represents different initial states, as 
          in Fig.~\ref{f:fig_moments_higher}. 
          Again, the rescaled data (right column) is well fitted by a Gaussian,
          $\mathrm{exp}\left[-(Jt/\tau)^2\right]$, with $\tau = 1.52$.
          }

 \label{f:dyn4local}
\end{figure}

For this, we first consider the following $4$-local operators defined in 
terms of the $\hat{\kappa}_{i}$s introduced in Eq.~(\ref{e:stagmagSYK}), 
\begin{equation}
 \hat{S}_j = - \hat{\kappa}_{2j-1}\hat{\kappa}_{2j}
                   = (\hat{n}_{4j-2} -\hat{n}_{4j-3})(\hat{n}_{4j-1}
                   -\hat{n}_{4j}).
 \label{e:4local}
\end{equation}
The system average of these operators is defined as 
$\hat{S} =\frac{1}{(N/4)}\sum_{j} \hat{S}_j$, where $j \in \{1,2,...,N/4\}$. 
For an $N=16$ system, we can construct four such $4$-local operators, i.e., 
$\hat{S}_1$, 
$\hat{S}_2$, $\hat{S}_3$, and $\hat{S}_4$. In 
Fig.~\ref{f:dyn4local}, we show the representative evolution of 
$S_1(t) = \langle\psi(t)|\hat{S}_1|\psi(t)\rangle$ as well as of 
$S(t) = \langle\psi(t)|\hat{S}|\psi(t)\rangle$. As in the cases of the QFI and 
moments of the staggered magnetization, we recover the super-exponential 
universal equilibration dynamics of these $4$-local operators.

\begin{figure}[ht!]
 \includegraphics[height = 6.0cm]{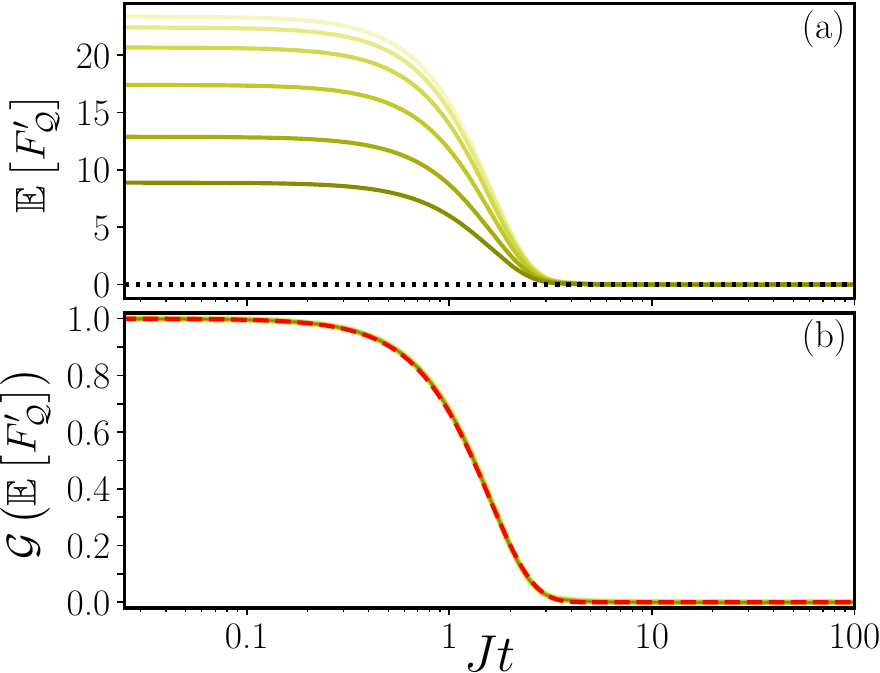}
 \caption{
	  Universal equilibration dynamics of the QFI, 
	  $F^{\prime}_{\mathcal{Q}}$, averaged over $2000$ disorder
          realizations of the complex 
	  SYK$_4$ Hamiltonian for $Q=6$ fermions occupying $N = 12$ fermionic modes.
      (a) QFI (yellow curves) with respect to the operator $\hat{T}$ 
      defined in Eq.~(\ref{e:spinflip}).
      Dark to light shading represents different initial states, as in 
      Fig.~\ref{f:fig_moments_higher}. 
      The dotted black line marks the QFI calculated with respect to the Gibbs 
      infinite temperature state.
      (b) Under rescaling with the function $\mathcal{G}$ as defined in
      Eq.~\eqref{e:rescaling}, the data collapse onto a single universal curve, 
      which is well fitted by a Gaussian, 
      $\mathrm{exp}\left[-(Jt/\tau)^2\right]$, with $\tau = 1.58$ (dashed red 
      curve).
      }
 \label{f:dynqfispin}
\end{figure}

All the operators considered so far are diagonal with respect to 
the Fock-space spanned by the occupation number basis vectors
$\{\ket{n_{1},n_{2},...,n_{N}}\}$. Here, we additionally investigate the dynamics of a 
non-diagonal operator, 
\begin{equation}
 \hat{T} = \sum_{i=1}^{N/2}(\hat{c}_{2i}^{\dagger}\hat{c}_{2i-1}
         + \mathrm{H.c.}) .
\label{e:spinflip}
\end{equation}
In Fig.~\ref{f:dynqfispin}, we present the dynamics of the QFI, 
$F^{\prime}_{\mathcal{Q}}$, computed with 
respect to this operator. Similar to all the previous cases, we again obtain 
the super-exponential universal equilibration dynamics of this observable.

\section{Details of ME derivation}\label{app:MEapp}

Here, we provide further details on the derivation of the Lindblad master 
equation (ME) presented in 
Sec.~\ref{s:master_eq}.
\paragraph{Functional derivative.---}
The functional derivative presented in Eq.~\eqref{e:funcderiv} is obtained from 
the 
integrated von~Neumann equation of a single Hamiltonian realization,
\begin{equation}
\hat{\rho}(t) = \hat{\rho}(0) - i\int_{0}^{t}dt_1 
\comm{\hat{H}(t_1)}{\hat{\rho}(t_1)},
\end{equation}
as
\begin{equation}\label{e:funcderiv_exact}
\begin{split}
\frac{ \var \hat{\rho}[\xi,t] }{ \var \xi^{(\alpha)}_{k_\alpha}(t') } =& 
-i 
\comm{ \hat{h}_{k_\alpha}^{(\alpha)} }{ \hat{\rho}(t') } \Theta(t-t')\\
&- i\int_{0}^{t}dt_1 
\comm{ \hat{H}(t_1) }{ \frac{\var \hat{\rho}[\xi,t_1] }{ \var 
\xi^{(\alpha)}_{k_\alpha}(t') 
} } 
\Theta(t-t') ,
\end{split}
\end{equation}
in which the step-function $\Theta$ arises from causality. The 
above recursive expression for the functional derivative yields 
a 
series of nested commutators, which to lowest order reduces to 
the first term of Eq.~\eqref{e:funcderiv_exact}. As mentioned in the main text, 
truncation to this lowest order is motivated by the fact that the resulting 
evolution equation [given by Eq.~\eqref{e:preMarkov}]
is formally equivalent to that obtained when making the decorrelation 
assumption in the study of stochastic evolution equations 
\cite{GardinerZoller_textbook,vanKampen_textbook}. 
This is readily seen in the interaction picture generated by     
$\hat{H}_0(t)$: 
Integrating the von~Neumann equation of an individual Hamiltonian realization 
one 
obtains a self-consistent integral equation for $\hat{\rho}(t)$.
By inserting this back into itself once, taking the disorder average and 
finally differentiating with respect to time, one finds an equation which 
 reduces to the interaction picture version of Eq.~\eqref{e:preMarkov} after 
performing the decorrelation approximation
$\eave{\xi^{(\alpha)}_{l_\alpha}(t)\xi^{(\beta)}_{k_\beta}(t') 
    \hat{\rho}(t')}\simeq\eave{\xi^{(\alpha)}_{l_\alpha}(t) 
    \xi^{(\beta)}_{k_\beta}(t')}\tilde{\rho}(t')$.

\paragraph{Lindblad form.---}
The final Lindblad master equation of Eq.~\eqref{e:me} is obtained from 
Eq.~\eqref{e:preMarkov} by first making the Markov approximation 
$\tilde{\rho}(t')\simeq \tilde{\rho}(t)$ and then expanding the double 
commutator. Simplifying 
our notation, this expansion is
\begin{equation}\label{e:commutator_expansion}
\sum_{l,k} f_{l,k}(t) (\hat{h}_l \hat{h}_k \tilde{\rho} - \hat{h}_l 
\tilde{\rho} \hat{h}_k  
- \hat{h}_k \tilde{\rho} \hat{h}_l + \tilde{\rho} \hat{h}_k \hat{h}_l),
\end{equation}
which is already reminiscent of a master equation in standard form. The latter 
is obtained in a final step in which we require the correlations to be 
symmetric under an exchange 
of the indices, i.e., $f_{l, k}(t) = f_{k, l}(t)$. This is trivially fulfilled 
for static processes such as those of the SYK model. For 
continuous processes, our requirement is equivalent to symmetry in 
time, i.e., $\eave{\xi_{l}(t) \xi_{k}(t')} = \eave{\xi_{l}(t') \xi_{k}(t)}$. We 
can then regroup the terms of 
Eq.~\eqref{e:commutator_expansion} as
\begin{equation}
\begin{split}
&\frac{1}{2}\sum_{l,k} \left[ f_{l,k}(t) (\hat{h}_l \hat{h}_k \tilde{\rho} - 
\hat{h}_l \tilde{\rho} \hat{h}_k  - \hat{h}_k \tilde{\rho} \hat{h}_l + 
\tilde{\rho} \hat{h}_k 
\hat{h}_l) + l \leftrightarrow k \right] \\
=& \sum_{k,l} 2 f_{k,l}(t) \left( \frac{1}{2}\acomm{\hat{h}_k 
    \hat{h}_l}{\tilde{\rho}}  - \hat{h}_l \tilde{\rho} \hat{h}_k \right) .
\end{split}
\end{equation}
We thus finally obtain the Lindblad master equation in non-diagonal form, 
given in the main text by Eqs.~\eqref{e:me}--\eqref{e:dissip}.

\paragraph{Cross-correlations of dissipation rates.---}
Here we comment on the origin of the cross-correlations for $\alpha=2,3$ 
which exist in the dissipation rates, defined in Eq.~\eqref{e:drates_syk4}, of 
the SYK$_4$ model.
To rewrite $\hat{H}_{\mathrm{SYK}_4}$ in the form of Eq.~\eqref{e:Hgeneric}, 
we partially order the indices as $i_1 > i_2, j_1 > j_2$ and use the 
anti-symmetry [see Eq.~\eqref{e:asymmetry}] of the SYK interactions.
We then find the three disorder terms given in Eq.~\eqref{e:syk4_xi}.
The point is now that for $\alpha=2,3$, correlations exist between pairs 
$\xi_{l_\alpha}^{(\alpha)} , \xi_{k_\alpha}^{(\alpha)} $ if multi-indices 
$l_\alpha, k_\alpha$ are mirror images, i.e., if $l_\alpha=i_1 i_2;j_1 j_2$ and 
$k_\alpha=j_1 j_2;i_1 i_2$.
This is because $\mathrm{Re} J_{i_1 i_2;j_1 j_2} = \mathrm{Re} J_{j_1 j_2;i_1 
i_2}$ and similarly $\mathrm{Im} J_{i_1 i_2;j_1 j_2} = -\mathrm{Im} J_{j_1 
j_2;i_1 i_2}$, due to Eq.~\eqref{e:asymmetry}. 

\paragraph{Limit of time-independent processes.---}
Here we explicitly show that the Lindblad ME of 
Eqs.~\eqref{e:me}--\eqref{e:Fint}, 
formally derived for Gaussian processes with arbitrary time-dependence, is 
valid in the limit of time-independent (static) processes 
$\xi_{k_\alpha}^{(\alpha)}(t) \to \xi_{k_\alpha}^{(\alpha)}$.

In this limit, the generic Hamiltonian defined in Eq.~\eqref{e:Hgeneric} and 
the corresponding 
ensemble averaged von~Neumann equation given by Eq.~\eqref{e:vNensave} formally 
remain the same, but now have time-independent disorder contributions
\begin{equation}\label{e:eavn_static}
\partial_t \tilde{\rho}(t) 
= -i  \comm{\hat{H}_0(t)}{ \tilde{\rho}(t) } 
-i\sum_{\alpha ,l_\alpha} \comm{ \hat{h}^{(\alpha)}_{l_\alpha}  }{ 
    \eave{\xi_{l_\alpha}^{(\alpha)} \hat{\rho}(t)}  } .
\end{equation}
We now have modified correlations $\eave{\xi_{l_\alpha}^{(\alpha)} 
\hat{\rho}(t)}$, 
in which $\hat{\rho}(t)$ is a function (as opposed to a functional) of the 
Gaussian 
distributed random numbers $\xi_{l_\alpha}^{(\alpha)}$ (as opposed to 
random functions). These 
correlations may, however, still be treated via the  
Novikov--Furutsu theorem, which simplifies accordingly 
to \cite{KlyatskinTatarskii72}
\begin{equation}\label{e:novikov_static}
\eave{\xi^{(\alpha)}_{l_\alpha} \hat{\rho}(\xi,t)} = \sum_{k_\alpha}     
F^{(\alpha)}_{l_\alpha,k_\alpha} 
\eave{\frac{d \hat{\rho}(\xi,t)}{d \xi^{(\alpha)}_{k_\alpha} }} ,
\end{equation}
where now we have disorder correlations 
\mbox{$F^{(\alpha)}_{l_\alpha,k_\alpha} = \eave{\xi^{(\alpha)}_{l_\alpha} 
\xi^{(\alpha)}_{k_\alpha}}$} with infinite correlation time.
This static version of the Novikov--Furutsu theorem is also known as 
Stein's lemma \cite{Liu94}.

The right-hand-side of Eq.~\eqref{e:novikov_static} now contains an ordinary 
total derivative (as opposed to a functional derivative), which we again obtain 
to lowest order from the integrated von~Neumann equation as
\begin{equation}\label{e:partialderiv}
\frac{d \hat{\rho}(\xi,t)}{d \xi^{(\alpha)}_{k_\alpha} } \simeq 
-i\int_{0}^{t}dt' 
\comm{\hat{h}^{(\alpha)}_{k_\alpha}}{\hat{\rho}(t')}.
\end{equation}
Substituting Eqs.~\eqref{e:partialderiv} and \eqref{e:novikov_static} into 
Eq.~\eqref{e:eavn_static} then yields
\begin{equation}\label{e:preMarkov_static}
\begin{split}
    \partial_t \tilde{\rho}(t)\! =&\! -i 
    \comm{\hat{H}_0(t)}{\tilde{\rho}(t)}\!\\
    &-\! \sum_{\alpha,l_\alpha, 
    k_\alpha}  F^{(\alpha)}_{l_\alpha,k_\alpha}
    \comm{ \hat{h}^{(\alpha)}_{l_\alpha} }{ \comm{ 
    \hat{h}^{(\alpha)}_{k_\alpha} } 
    { \int_{0}^{t}dt' \tilde{\rho}(t')} } .
\end{split}
\end{equation}
This is exactly the evolution equation that one obtains by taking the limit of 
time-independent processes, i.e., of infinite correlation times in 
Eq.~\eqref{e:preMarkov}. Taking the Markov approximation in 
Eq.~\eqref{e:preMarkov_static} thus yields the Lindblad master equation of 
Eqs.~\eqref{e:me}--\eqref{e:Fint} for time-independent correlations $ 
F^{(\alpha)}_{l_\alpha,k_\alpha}$. This shows that starting from time-dependent 
noise and then taking the limit of infinite correlation time is equivalent to 
working with static noise all the way. The advantage of the time-dependent 
formulation used in the main text is that it naturally incorporates static 
disorder and temporally fluctuating noise on the same footing.

\paragraph{Matrix representation of Liouvillian superoperator.---}
We numerically obtain the spectrum $\{\lambda_i\}$ and eigenmodes 
$\{\hat{\rho}_i \}$ of a Liouvillian superoperator $\mathcal{L}$ by numerically 
diagonalizing a matrix representation thereof. This 
representation is obtained by mapping $\mathcal{L} 
\to \sopvec{\mathcal{L} }$, with $\sopvec{\mathcal{L} }$ an operator acting 
on the duplicated 
Hilbert space $\mathcal{H} \otimes \mathcal{H}$ (see for instance 
Ref.~\cite{Minganti18}). 
Under this map, a density matrix $\hat{\rho}$ becomes a 
vector $\ket{\vec{\rho}}\in \mathcal{H} \otimes \mathcal{H}$ obtained by 
stacking the columns of the matrix representation of
$\hat{\rho}$. Left and right multiplication with operators transform as 
$\hat{A} \hat{\rho} 
\hat{B} \to \hat{B}^\intercal \otimes \hat{A} \ket{\vec{\rho}}$, where 
$\hat{B}^\intercal$ denotes the 
transpose of $\hat{B}$. 
For our general Liouvillian [see Eq.~\eqref{e:me}], we then have the matrix 
representation 
\begin{equation}\label{e:Lvec}
\begin{split}
\sopvec{\mathcal{L} }(t) =& -i[\mathds{1}\otimes \hat{H}_0(t) - 
(\hat{H}_0(t))^\intercal 
\otimes 
\mathds{1} ] + \sopvec{\mathcal{D}}, \\
\text{with } \sopvec{\mathcal{D}} = &\sum_\alpha \sum_{l_{\alpha}, k_{\alpha}} 
2 f^{(\alpha)}_{l_\alpha, k_\alpha}(t) \Bigl[ \left( 
\hat{h}^{(\alpha)}_{k_\alpha} \right)^\intercal \otimes  
\hat{h}^{(\alpha)}_{l_\alpha} \\
&- \frac{1}{2} \mathds{1} \otimes \hat{h}^{(\alpha)}_{k_\alpha} 
\hat{h}^{(\alpha)}_{l_\alpha} -\frac{1}{2} \left(\hat{h}^{(\alpha)}_{k_\alpha} 
\hat{h}^{(\alpha)}_{l_\alpha} \right)^\intercal \otimes \mathds{1}  \Bigr] .
\end{split}
\end{equation}

We obtain the matrix representation of the SYK$_4$ Liouvillian 
$\sopvec{\mathcal{L} }(t) = 
2t\sopvec{\mathcal{D}}$ (factoring out $2t$ as in the main text) by 
setting $\hat{H}_0=0$ and inserting 
Eqs.~\eqref{e:jops_syk4}--\eqref{e:drates_syk4} into Eq.~\eqref{e:Lvec}.

\section{Details on numeric simulation of exact dynamics}
Here, we provide a brief description of the algorithm implemented to solve the exact dynamics reported in this manuscript.
We exploit the particle number conservation of the SYK Hamiltonian $\hat{H}_{\mathrm{SYK}_{4}}$ [see Eq.~\eqref{e:Hsyk4}] by restricting our simulations to a given particle number sector of the Hilbert space.
In particular, all simulations reported in this manuscript are performed within the half-filling sector where the fermion number is $Q=N/2$, and the Hilbert space dimension is $D=N!/((N/2)!)^2$ (see Sec.~\ref{s:quench_protocol}).

The matrix representation of $\hat{H}_{\mathrm{SYK}_{4} }$ is constructed with respect to the fermion mode occupation number Fock basis of the $Q=N/2$ sector:
Each Fock state $\ket{s_a}$, where $a=1,\ldots, D$, is represented by one of the $N$-bit strings $s_a$ of which $Q$ bits are $1$ in order to represent the occupied fermion modes.
The Hamiltonian's matrix elements $H_{ab}\equiv \matrixel{s_a}{\hat{H}_{\mathrm{SYK}_{4}} }{s_b} $ are non-zero only for those pairs of states whose bit string representation have a Hamming distance $d(s_a,s_b)=0,2,4$.
This follows from the quartic operators $\hat{c}^{\dagger}_{i_{1}}\hat{c}^{\dagger}_{i_{2}} \hat{c}_{j_{1}}\hat{c}_{j_{2}}$ appearing in $\hat{H}_{\mathrm{SYK}_{4}} $, where $d(s_a,s_b)=0$ corresponds to $(i_1,i_2)=(j_1,j_2)$, $d(s_a,s_b)=2$ to $i_k=j_l\, ,i_{k'}\neq j_{l'}$ for $k,k',l,l'=1,2$, and $d(s_a,s_b)=4$ to $i_1\neq i_2 \neq j_1 \neq j_2$.
These non-zero elements are populated by independent random complex Gaussian variables $J_{i_{1}i_{2};j_{1}j_{2}}$, in accordance with Eqs.~\eqref{e:syk_statistics}~and~\eqref{e:asymmetry} to ensure Hermiticity of the Hamiltonian and antisymmetry of the interaction amplitudes under permutation of the indices.
In this way, a single realization of the random SYK$_4$ Hamiltonian is constructed.

For system sizes $N\leq 14$, $D$ is sufficiently small such that the full eigensystem (energy basis) of the above matrix representation can be solved exactly.
For this we employ diagonalization routines from \verb|LAPACK| 
\cite{lapack99}.
Time evolution of a given initial state $\ket{\psi(0)}$ is then solved by rotating the corresponding vector representation into the SYK energy basis and calculating the time-dependent phases $\exp(-i\epsilon_n t)$ for any time $t\in \RR$, where $\epsilon_n$ for $n=1,\ldots, D$ are the eigenenergies of $\hat{H}_{\mathrm{SYK}_{4}}$.

For system sizes $N>14$, $D$ is so large as to prohibit the above exact numeric solution of the entire eigensystem.
Instead, we utilize a Runge--Kutta $4$ (RK$4$) method to solve the Schr\"{o}dinger equation for $\hat{H}_{\mathrm{SYK}_{4} }$.
The matrix-vector multiplication employed within the RK$4$ method makes use of a sparse matrix representation of $\hat{H}_{\mathrm{SYK}_{4} }$ in order to exploit the large amount of zero matrix elements, and thus enhance the speed of the algorithm.
Further reduction of computation time is achieved by parallelizing this matrix-vector multiplication via \verb|MPI| 
methods \cite{mpi40}, allowing us to exactly solve (within numeric precision) the dynamics for system sizes up to $N=20$ at $Q=N/2$.
To ensure accuracy of this RK$4$ based algorithm, we benchmark its dynamics against those of the above exact diagonalization scheme for $N\leq 14$, verifying that they agree.

Finally, we average over the dynamics of multiple disorder realizations of $\hat{H}_{\mathrm{SYK}_{4} }$.
To this end, we again utilize \verb+MPI+ 
methods to solve the dynamics of multiple disorder realizations in parallel.

\bibliographystyle{myunsrtnat}
\bibliography{syk}


\end{document}